%% file: paper.tex
\documentclass[preprint,12pt, 1p, review]{elsarticle}

\usepackage{wrapfig}
\usepackage{subfigure}
\usepackage{caption}
\usepackage{makecell}
\usepackage{float}
\usepackage{hyperref}

\usepackage{amssymb}
\usepackage{amsmath}
\usepackage{bm}
\usepackage{mathtools}
\DeclarePairedDelimiter{\ceil}{\lceil}{\rceil}
\usepackage{algorithm}
\usepackage[noend]{algpseudocode}
\usepackage{color}
\usepackage{csvsimple}
\usepackage{setspace}
\usepackage[font={scriptsize, stretch=1.},skip=0pt]{caption}

\usepackage{lineno}

\journal{Journal of Computational Physics}

\newcommand{\R}{\mathbb{R}}
\newcommand{\x}{\mathbf{x}}
\newcommand{\y}{\mathbf{y}}
\newcommand{\z}{\mathbf{z}}
\newcommand{\X}{\mathbf{X}}

\newcommand{\W}{\mathbf{W}}
\newcommand{\bi}{\mathbf{b}}
\newcommand{\ax}{\mathbf{a}}

\newcommand{\calG}{\mathcal{G}}
\newcommand{\ux}{\mathbf{u}}

\newcommand{\E}{\mathbb{E}}
\newcommand{\btheta}{\bm{\theta}}
\newcommand{\bxi}{\bm{\xi}}
\newcommand{\bzeta}{\bm{\zeta}}
\newcommand{\bTheta}{\bm{\Theta}}
\newcommand{\calD}{\mathcal{D}}
\newcommand{\calL}{\mathcal{L}}
\newcommand{\calN}{\mathcal{N}}
\newcommand{\calR}{\mathcal{R}}
\newcommand{\calS}{\mathcal{S}}

\begin{document}
\begin{frontmatter}
\title{Deep UQ: Learning deep neural network surrogate models for high dimensional uncertainty quantification.}

\author{Rohit K. Tripathy}
\ead{rtripath@purdue.edu}
\author{Ilias Bilionis\corref{cor1}}
\ead{ibilion@purdue.edu}
\ead[url]{https://www.predictivesciencelab.org/}
\address{Predictive Science Laboratory, School of Mechanical Engineering, \\
585 Purdue Mall, Purdue University, West Lafayette, IN 47907-2088, USA}
\cortext[cor1]{Corresponding author}

\input{abstract.tex}
\end{frontmatter}



\input{introduction.tex}

\input{methodology.tex}

\input{examples.tex}

\input{conclusion.tex}

\bibliographystyle{model1-num-names}
\bibliography{bibliography.bib}
\end{document}

%% file: abstract.tex
\begin{abstract}
 State-of-the-art computer codes for simulating real physical systems are often characterized by vast number of input parameters. Performing uncertainty quantification (UQ) tasks with Monte Carlo (MC) methods is almost always infeasible because of the need to perform hundreds of thousands or even millions of forward model evaluations in order to obtain convergent statistics. One, thus, tries to construct a cheap-to-evaluate surrogate model to replace the forward model solver. For systems with large numbers of input parameters, one has to deal with the curse of dimensionality - the exponential increase in the volume of the input space, as the number of parameters increases linearly. Suitable dimensionality reduction techniques are used to address the curse of dimensionality. A popular class of dimensionality reduction methods are those that attempt to recover a low dimensional representation of the high dimensional feature space. However, such methods often tend to overestimate the intrinsic dimensionality of the input feature space. In this work, we demonstrate the use of deep neural networks (DNN) to construct surrogate models for numerical simulators. We parameterize the structure of the DNN in a manner that lends the DNN surrogate the interpretation of recovering a low dimensional nonlinear manifold. The model response is a parameterized nonlinear function of the low dimensional projections of the input. We think of this low dimensional manifold as a nonlinear generalization of the notion of the \textit{active subspace}. Our approach is demonstrated with a problem on uncertainty propagation in a stochastic elliptic partial differential equation (SPDE) with uncertain diffusion coefficient. We deviate from traditional formulations of the SPDE problem by not imposing a specific covariance structure on the random diffusion coefficient. Instead we attempt to solve a more challenging problem of learning a map between an arbitrary snapshot of the diffusion field and the response.   
\end{abstract}

%% file: introduction.tex
\section{Introduction}
\label{sec:intro}
With the tremendous increase in the availability of computational resources, computer codes which simulate physical systems have become highly sophisticated. Today, state-of-the-art numerical simulators are parameterized by a very large number of quantities which are used to describe material properties, initial conditions, boundary conditions, constitutive laws, etc. It is often the case, that many of the inputs to a numerical simulator are not known exactly. This raises the question - how defensible are the predictions from such numerical simulators? How do we objectively assess the effects of the uncertainties in model inputs on the quality of the model predictions? Answering such questions are at the core of the field of uncertainty quantification (UQ) \cite{smith2013, sullivan2015}. Specifically, the task of rigorously quantifying the effect of input parameter uncertainty on the model outputs is known as the forward UQ or uncertainty propagation (UP) problem. 

The simplest method for tackling the UP problem is the Monte Carlo (MC) method \cite{liu2008,mooney1997,robert2004}. The basic idea of MC is that one can compute empirical estimates of the statistics of some quantity of interest (QoI) by sampling averages. The MC method is guaranteed to converge in the limit of infinite samples. MC methods, and its advanced variants, are routinely applied to UQ tasks such as UP \cite{baraldi2008}, inverse problems \cite{mosegaard1995, mosegaard2002}, model calibration \cite{bilionis2015} and stochastic optimization \cite{spall2005}. The computational time to convergence of MC methods is independent of the number of the stochastic dimensions. However, the number of samples needed by MC methods, to obtain convergent statistics is large, typically being of the order of hundreds of thousands or millions. This makes MC methods unsuitable for UQ tasks involving expensive computer codes. 

We typically deal with expensive computer codes, by building a cheap-to-evaluate surrogate of the response surface. To do this, a set of locations in the uncertain parameter-space are carefully selected and the forward model is evaluated at these locations. This produces a set of independent observations of the model response. The total number of such simulations to be performed is determined by one's computational budget and desired accuracy. Because the surrogate model can be queried very cheaply, one can use it as a replacement of the original simulator and perform UQ tasks using MC techniques.  Popular choices for surrogate models in the literature include, Gaussian processes \cite{rasmussen2006gaussian, tripathy2016gaussian, bilionis2012multi, bilionis2013multi, chen2015uncertainty}, polynomial chaos expansions \cite{xiu2002wiener, xiu2003modeling, xiu2002modeling,najm2009uncertainty}, radial basis functions \cite{park1991universal} and relevance vector machines \cite{bilionis2012multidimensional}. Despite their success, these methods become intractable for problems in which the number of input stochastic dimensions is large.

In order to construct a surrogate response surface for a multivariate function with a large number of uncertain parameters, one has to overcome the phenomenon known as the \textit{curse of dimensionality}, a term coined by the mathematician Richard Bellman \cite{keogh2011curse}. It describes the exponential increase in the volume of a function input space even as the space dimensionality increases linearly. The implication of the curse of dimensionality is that to sufficiently explore a high dimensional space, one must visit an exponentially large number of points in that space. As a concrete example, suppose the task of approximating a surrogate model for a univariate function can be done by visiting 10 locations in the input space and evaluating the forward model at those input locations. For a bivariate function of similar lengthscale, one would need to visit roughly $10 \times 10 = 100$ points in the input space to maintain a similar level of accuracy of the constructed surrogate. Generalizing, a $d$-variate function requires visiting $\mathcal{O}(10^d)$ locations in the input space and evaluating the forward solver at those locations. Even if the forward model is inexpensive to evaluate, attempting to naively construct surrogate response surfaces for high dimensional functions is a futile task.

Suitable \textit{dimensionality reduction} (also known as \textit{model order reduction}) techniques have to be employed in order to deal with the curse of dimensionality. 
The simplest way of doing so is to rank the input dimensions in order of their ``importance", and rejecting those input dimensions which contribute the least to the outcome of the numerical simulation. Techniques that adopt this approach include sensitivity analysis \cite{saltelli2000sensitivity} and automatic relevance determination (ARD) \cite{neal1998assessing}. These methods, while effective for problems with a small number of uncorrelated input dimensions, are not useful for problems involving functional uncertainties, such as stochastic partial differential equations (SPDE). 

Many common dimensionality reduction techniques follow  a simple idea : project the high dimensional inputs, onto a low-dimensional subspace which captures most of the information contained in the original input. 
In the UQ community, the most common dimensionality reduction method used is the \textit{Karhunen-Lo\`eve expansion} (KLE) \cite{ghanem2003stochastic, spanos1989stochastic}. 
The KLE involves computing an eigendecomposition of the covariance function associated with the uncertain functional input. 
The eigenfunctions represent a set of orthogonal basis functions and the decay of the eigenvalues determine the set of basis functions to be retained for the purpose of constructing a low dimensional approximation of the high dimensional random field. In the machine learning (ML) community, this is more popularly known as the \textit{principal component analysis} (PCA) \cite{jolliffe2002principal, wold1987principal}, whose goal is to produce a low-rank approximation of the empirical centered covariance matrix of the available input data. The result of such a computation is an orthogonal matrix, which maps a point in the high dimensional input space to a point on a low dimensional manifold, such that there is minimal reconstruction error. The obvious drawback of PCA is the fact that one is constrained to discover only linear projections. Furthermore, PCA is an unsupervised technique, which means that it only looks at samples of the input and does not consider information contained in the model outputs. As a result, PCA tends to overestimate the intrinsic dimensionality of the system. Thus, inspite of a reduction in the total number of input dimensions, the reduced representation remains very high dimensional and thereby unsuitable for surrogate model construction. One can alleviate the linearity  constraint by using the kernel PCA (KPCA) \cite{scholkopf1997kernel, ma2011kernel}, which uses the kernel trick to discover nonlinear manifolds. Nevertheless, KPCA is also an unsupervised technique that ignores model output information. 

A recent advancement in dimensionality reduction is \textit{active subspaces} (AS) \cite{constantine2014active, lukaczyk2014active, constantine2014computing, constantine2015exploiting, constantine2015discovering, bauerheim2014uncertainty}. An active subspace is a low dimensional linear manifold embedded in the input space which captures most of the model output variation. It does so by recovering an orthogonal projection matrix obtained through an eigendecomposition of an empirical covariance matrix of the model output gradients. In the absence of model output gradients (a scenario typical in engineering applications), one can estimate the projection matrix by posing it as a hyperparameter in a Gaussian process regression model and learning it from the available data through a maximum likelihood (MLE) computation \cite{tripathy2016gaussian}. While the upshot of the AS method is that one utilizes the information contained in the model outputs along with the model inputs, one is still constrained to discover only linear manifolds of the data. 

In this work, we propose a systematic approach for constructing surrogate models using \textit{deep neural networks} (DNNs) \cite{bishop1995neural, goodfellow2016deep, schmidhuber2015deep, haykin2009neural}. Neural networks (NNs) (or artificial neural networks (ANNs)) are a class of function approximators that have shot to prominence in recent years because of breakthrough successes achieved in numerous artificial intelligence (AI) tasks such as image classification \cite{krizhevsky2012imagenet, wan2013regularization, graham2014fractional, clevert2015fast, lee2016generalizing} and autonomous driving \cite{huval2015empirical, chen2015deepdriving}. The idea of DNNs is not new. The reason for their increased usage and popularity in recent times is due to:   1. Advancements in computer hardware leading to widespread availability of graphics processing units (GPUs) for accelerated computation; 2. Advances in stochastic optimization including techniques such as \textit{Adam} \cite{kingma2014adam}, \textit{RMSprop} \cite{tieleman2012lecture}, \textit{AdaGrad} \cite{duchi2011adaptive}, \textit{AdaDelta} \cite{zeiler2012adadelta} etc.; 3.  Regularization techniques such as  \textit{dropout} \cite{srivastava2014dropout}; and, 4. Development of easy-to-use software libraries, such as \textit{Tensorflow} \cite{abadi2016tensorflow}, \textit{PyTorch} \cite{paszke2017pytorch} and \textit{Theano} \cite{bergstra2010theano}. 
\\
The basic idea of DNNs is that one can represent multivariate functions through a hierarchy of features of increasing complexity. The most typical example of a DNN is a feedforward multilayer perceptron (MLP). A highly attractive property of MLPs is that, under mild assumptions on the underlying function being approximated, they are universal approximators \cite{hornik1989multilayer}. This means that any continuous function, regardless of its complexity, can be approximated with a neural network of just one layer with a sufficient number of hidden units. DNNs tackle the curse of dimensionality through a series of nonlinear projections of the input into exploitable latent spaces. In fact, PCA can be thought of as a special case of a DNN with no hidden layers such that the latent space is recovered through an orthogonal projection of the input. 

The powerful nonlinear function approximation capabilities coupled with the scalability of DNNs to high dimensions offers a very promising direction of research for the UQ community, with the potential to significantly improve upon state-of-the-art capabilities. \cite{khoo2017solving} use a 3 layer convolutional DNN to learn a map between input coefficients of SPDEs to a functional of the PDE solution.  \cite{zhu2018bayesian} use a Bayesian fully convolutional encoder-decoder network to solve an image-to-image regression task mapping the random input coefficient field of an elliptic SPDE to the corresponding solution. These papers offer encouraging results for challenging problems in UQ. However, they are only applicable to tasks where input parameters and output quantities can be treated as images and they require a lot of cross validation to learn an optimal network structure. Specifically in the context of SPDEs, we are interested in learning a surrogate that can make predictions at spatial locations other than those included in the discretization of the underlying deterministic numerical solver. 

The task of selecting the architecture of the network and values for hyperparameters such as regularization constants is a persistant problem in the application of DNNs. Under constraints of limited data, this task assumes added importance. In this work, we present a methodology based on MLPs where we parameterize our network in a way that lends it the interpretation of discovering a low dimensional nonlinear manifold that captures maximal variation of the model outputs. We think of this procedure as discovering a nonlinear active subspace. The \textit{projection function}, which connects the high dimensional input, to the low dimensional manifold, is linked to the scalar model output through a linear transformation. We utilize a combination of Bayesian global optimization (BGO) \cite{brochu2010tutorial} and grid search to select the best setting of the network hyperparameters and determine the appropriate structure. 

This paper is organized as follows. In Sec. \ref{sec:methodology}, we discuss the general setup of the problem we are dealing with. Sec. \ref{sec:up} provides a mathematical description of the UP problem. In Sec. \ref{sec:surr_mod}, we discuss the task of surrogate modeling. In Sec. \ref{sec:dnn} through Sec. \ref{sec:grad_opt}, we discuss the process of constructing a DNN surrogate model, including the parameterization of the network architecture and the optimization of the network parameters. We conclude Sec. \ref{sec:methodology} with a description of the procedure we use to select network hyperparameters. In Sec. \ref{sec:examples}, we demonstrate our methodology on a SPDE with uncertain diffusion coefficient. A novelty of our work is that we do not make any assumption on the regularity and lengthscales of the uncertain diffusion. Specifically, we construct a surrogate of the SPDE solver which can accurately predict the response when tested with input random fields that may not be structurally similar (in terms of smoothness and lengthscales) to samples of the input in the training dataset. This deviates from the standard formulation of this problem in the UQ literature, where a specific covariance structure is imposed on the uncertain parameter. As a result, our problem is not amenable to the application of preliminary dimensionality reduction using the KLE, thereby making it far more challenging than the traditional formulation of the problem. We wrap up this article with concluding remarks in Sec. \ref{sec:conclusion}.

%% file: methodology.tex
\section{Methodology}
\label{sec:methodology}
Suppose we have a computer code simulating a physical phenomena. Mathematically, we represent this simulator as a function $f:\mathcal{X} \rightarrow \mathcal{Y}$. $f$ accepts a vector of inputs $\bxi \in \mathcal{X} \subseteq \R^D$ where $\bxi$ could specify material properties, external loads, boundary conditions, initial conditions, etc. The output of the computer code is some scalar quantity of interest $y = f(\bxi) \in \mathcal{Y} \subseteq \R$. Typically, $f$ depends on the solution of some PDE which depends on $\bxi$. Furthermore, $f$ is unknown in closed form and information about it can only be obtained by querying the simulator at feasible values of $\bxi$. We allow for the possibility that the output observation, $y$, may be a noisy estimate of the true solution $f(\bxi)$, i.e., $y = f(\bxi) + \epsilon$, where $\epsilon$ is Gaussian noise. Finally, the dimensionality, $D$, of the input vector $\bxi$ is large, potentially of the order of hundreds or thousands. Given a finite number of evaluations of the simulator, the task of constructing a surrogate function, $\hat{f}$, for the true response surface $f$ becomes computationally infeasible without resorting to dimensionality reduction. 

\subsection{Uncertainty propagation}
\label{sec:up}
Suppose the inputs $\bxi$ to the function $f$ are not known exactly (a common scenario in numerous engineering tasks). We formalize our beliefs about $\bxi$ using a suitable probability distribution:
\begin{equation}
\label{eqn:input_uncertainty}
\bxi \sim p(\bxi).
\end{equation}

Given our beliefs about $\bxi$, we wish to characterize the statistical properties of the output $f(\bxi)$ such as the mean:
\begin{equation}
\label{eqn:out_mean}
\mu_f = \int f(\bxi) p(\bxi) \mathrm{d}\bxi,
\end{equation}
the variance, 
\begin{equation}
\label{eqn:out_var}
\sigma_{f}^{2} = \int \big(f(\bxi) - \mu_f\big)p(\bxi)\mathrm{d}\bxi, 
\end{equation}
and the probability density, 
\begin{equation}
\label{eqn:out_pdf}
p_f(y) = \int \delta \big( y-f(\bxi) \big) p(\bxi) \mathrm{d}\bxi.
\end{equation}
This is formally known as the uncertainty propagation problem (UP). 

\subsection{Surrogate model structure}
\label{sec:surr_mod}
Since the function $f$ is not known in closed form we resort to numerical approximations of the integrals in Eq. (\ref{eqn:out_mean}), (\ref{eqn:out_var}) and (\ref{eqn:out_pdf}). The easiest way to do so, would be to use the MC method. Unfortunately, as discussed earlier in Sec.(\ref{sec:intro}), the MC method is computationally infeasible because of slow convergence in the number of forward model evaluations. Furthermore, information about $f$ can only be obtained by querying the computer code at carefully selected design points. Say, we have $N$ design points, which we collectively denote as $\X$:
\begin{equation}
\label{eqn:design_points}
\X = ( \bxi_1, \bxi_2, \cdots, \bxi_N ).
\end{equation}
$\X$ is an $N \times D$ matrix, with each row representing a single sample from $p(\bxi)$. We evaluate the computer code for each of these $N$ samples and obtain a potentially noisy estimate $y_i = f(\bxi_i) + \epsilon$ of the model output, $\forall i \in \{1, 2, \cdots, N\}$. We collectively represent all samples of the model output as, 
\begin{equation}
\label{eqn:model_out}
\mathbf{y} = ( y_1, y_2, \cdots, y_N ). 
\end{equation}
$\mathbf{y}$ is an vector in $\R^N$, with each element of the vector representing a sample of the output. We denote the inputs and the outputs taken together as $\mathcal{D} = ( \mathbf{X}, \mathbf{y})$. Thus, the task of building a surrogate model can be summarized as follows - given data $\mathcal{D}$ collected by querying the computer code at a finite number of input locations, we wish to build an approximation $\hat{f}$ of the true model $f$. We propose a form of the surrogate model $\hat{f}$, which projects the input data onto a nonlinear low dimensional manifold, i.e., $\hat{f}:\mathcal{X} \rightarrow \R$, such that,
\begin{equation}
\label{eqn:dnn_as}
\hat{f}(\bxi) = g(\bzeta) = g(h(\bxi)),
\end{equation}
where, $\bzeta \in \mathcal{Z} \subseteq \R^d$ is the projected input corresponding to the true input $\x$. We call the function $h:\mathcal{X} \rightarrow \mathcal{Z}$, the \textit{projection function} and the function $g:\mathcal{Z} \rightarrow \mathcal{Y}$, the \textit{link function}. The link function is a generic nonlinear function of the projected inputs, $\bzeta$. 
One immediately recognizes this structure as a generalization of the active subspace surrogate in \cite{constantine2014active,tripathy2016gaussian} which expresses $\hat{f}$ as:
\begin{equation}
\label{eqn:lin_as}
\hat{f} = g(\mathbf{W}^T \bxi).
\end{equation}
The proposed structure in Eq. \ref{eqn:dnn_as} is capable of capturing directions which explain most of the variation in the model output. Alternatively, one also recognizes the above construction of the projection function as being the encoder section of neural network autoencoders\footnote{An autoencoder is a kind of DNN used to recover a low dimensional embedding of a high dimensional space.} \cite{goodfellow2016deep}. The complete structure is posed as a Deep Neural Network (DNN) which we describe in the following section. 

\subsection{Structure of a feedforward Deep neural network}
\label{sec:dnn}
\begin{figure}[h]
\subfigure[]{
\includegraphics[width=0.45\textwidth]{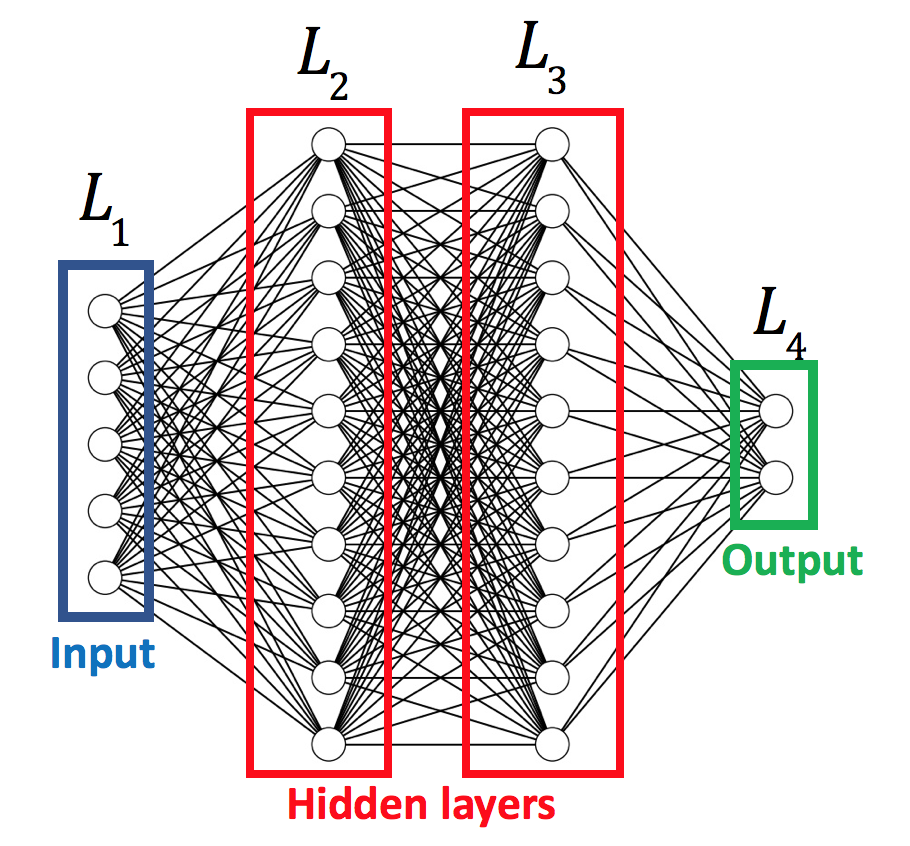}
\label{fig:nn_schematic}
}
\subfigure[]{
\includegraphics[width=0.45\textwidth]{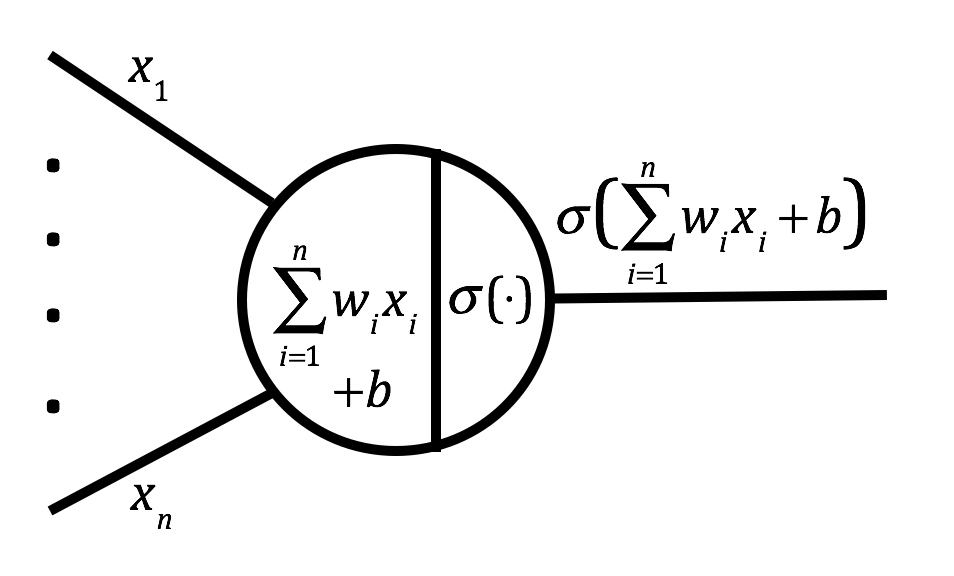}
\label{fig:neuron_schematic}
}
\caption{\ref{fig:nn_schematic}-Schematic of a neural network (NN). \ref{fig:neuron_schematic} - Schematic of a single neuron.}
\label{fig:schematic}
\end{figure}

Neural networks (NN) are a powerful class of data-driven function approximation algorithms which represent information through a hierarchy of features. Each step in the hierarchy, beginning with the input, and ending with the final output, is known as a \textit{layer}. Intermediate layers are known as \textit{hidden layers}. By manipulating the number of hidden layers and the size of each hidden layer, one can learn functions of arbitrary complexity. The sizes of the input layer and output layer are fixed and determined by the dimensionality of the input and output. Fig. \ref{fig:nn_schematic} shows a schematic of a NN with 2 hidden layers. Each circle in the schematic of the NN represents the fundamental unit of computation in a NN, known as the \textit{neuron}. A neuron accepts one or more inputs and produces an output by performing a linear transformation followed by an element-wise nonlinear transformation. A schematic of a single neuron and the computations taking place within it are shown in Fig. \ref{fig:neuron_schematic}. We discuss the symbols in full detail in the proceeding paragraphs. 

A DNN is simply a NN with a large number of hidden layers. The output of a layer is known as the \textit{activation}. The activation from one layer of a DNN is used as the input to the next layer. The activation produced by the $j^{th}$ hidden layer of the DNN is given by:

\begin{equation}
\label{eqn:lj_act}
\z^{(j)} = \sigma(\W^{(j)}\z^{(j-1)} + \bi^{(j)}),\ \forall j \in \{1, 2, \cdots, L\},
\end{equation}
where $\W^{(j)} \in \R^{d_j \times d_{j-1}}$, $\bi^{(j)} \in \R^{d_{j}}$ and 
$d_{j}$ is the number of neurons in the $j^{th}$ hidden layer. Note that $\z^{0}$ is the input $\bxi$ and $d_{0}= D$.
$\sigma$ is a nonlinear function applied element-wise on its arguments. Popular choices for $\sigma$ include the \textit{logistic} function, the \textit{hyperbolic tangent} function or the \textit{rectified linear unit} (or ReLU) function \cite{goodfellow2016deep}. The ReLU function has been utilized extensively in recent times, and has been shown to eliminate the need for an unsupervised pretraining phase while training deep architectures \cite{glorot2011deep}. 
\begin{wrapfigure}{r}{0.6\textwidth}
\includegraphics[width=0.6\textwidth]{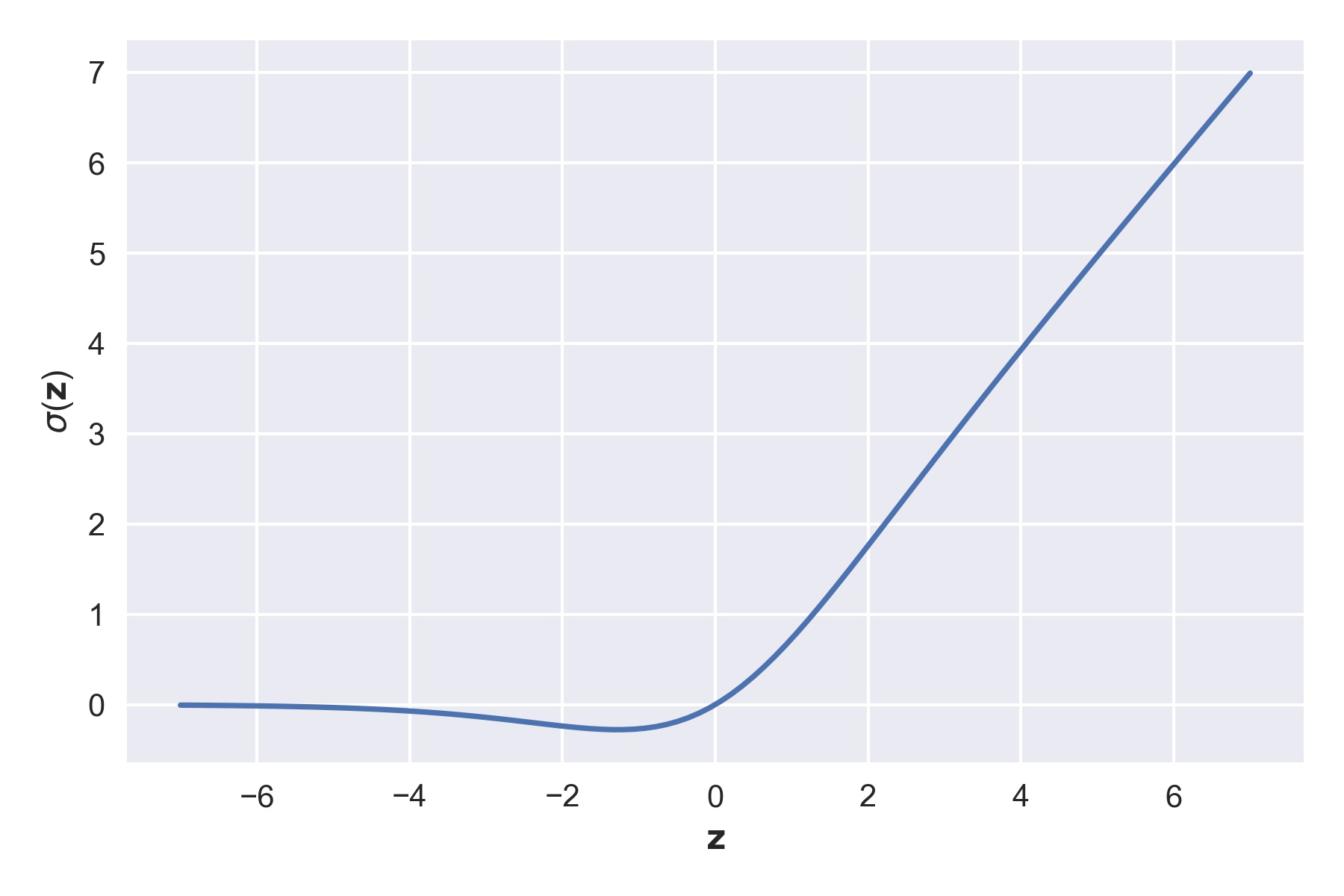}
\caption{Swish activation with $\gamma=1$.}
\label{fig:swish_act_func}
\end{wrapfigure}
However, a recent result described in \cite{ramachandran2017searching} demonstrates the superior performance of the \textit{Swish} activation function defined as follows: 
\begin{equation}
\label{eqn:swish_act_func}
\sigma(\z) = \frac{\z}{1 + \exp(-\gamma\z)},
\end{equation}
such that $\gamma$ is either a constant or a hyperparameter to be learned from data. In this work, we use the Swish activation function with $\gamma = 1$. 

The quantities $\W^{(j)}$ and $\bi^{(j)}$, $\forall j \in \{1, 2, \cdots, L+1\}$, are known as the \textit{weights} and \textit{biases} of the  network, respectively. Collectively, they are known as the \textit{parameters} of the network, $\btheta = \{\W^{(j)}, \bi^{(j)}\}_{j=1}^{L+1} \in \bTheta$. The weights and biases together, fully describe the structure of the network, known as the \textit{network architecture}. 

\subsection{Training a deep neural network}
As discussed in the previous section, $\hat{f}$ is a parameterized function with parameters $\btheta$. Estimating $\btheta$ reduces to the problem of minimizing
a loss function $\calL(\btheta; f)$, which captures the mismatch between $f$ and $\hat{f}$. For regression tasks $\calL$ is typically chosen to be the mean squared error. In practice, we do not have access to the function $f$; only a limited set of observations, $\calD$. 
Suppose $\calD$ is a dataset of $N$ examples, with the $i^{th}$ example denoted as $\calD_i = (\x_i, y_i)$.
The training problem is cast as minimizing the mismatch between a prediction $\hat{f}(\x_i)$ and the correct output, $y_i$. 
\begin{equation}
\label{eqn:train_dnn}
\btheta^* = \underset{\btheta}{\arg\min} \frac{1}{N} \sum_{i=1}^{N} \calL(\btheta;y_i).
\end{equation}

In practice, the averaging in Eq. \ref{eqn:train_dnn} is performed over a small randomly sampled subset (or \textit{mini-batch}) $\calD_M \subset \calD$, at each iteration of the optimization procedure. 

\subsection{Regularized loss function}
Recall that the output of the computer code at a given input location $\x$ is a potentially noisy estimate of $f(\x)$. Under the Gaussian noise assumption, the likelihood model is given by:
\begin{equation}
\label{eqn:likelihood}
p(y| \bxi, \btheta) = \calN(y| \hat{f}(\bxi;\btheta), \sigma^2).
\end{equation}
The unknown variance, $\sigma^2$, captures the discrepancy due to all sources of error including discretization errors, model discrepancy, noise, etc. 
It is easy to see that maximizing the logarithm of the conditional likelihood $p(y|\bxi, \btheta)$ is equivalent to minimizing the mean squared error of the examples in the training dataset.
Since DNNs are prone to overfitting \cite{srivastava2014dropout}, one resorts to penalizing the misfit function with an appropriate penalty term known as a regularizer. This ensures that the DNN generalizes better to unseen data. Popular choices for regularization include the scaled $L_1$ norm or $L_2$ norm of the weights \cite{goodfellow2016deep}. The $L_1$ norm penalty is known to promote sparsity in the estimator $\btheta^*$. On the other hand, the $L_2$ norm penalty drives the values of the weights close to 0. The \textit{elastic net} regularizer introduced in \cite{zou2005regularization} is a mixture between the $L_1$ and $L_2$ regularizers and is known to combine the advantages of $L_1$ and $L_2$ penalties (See \cite{zou2005regularization}). While typically the $L_1$ and $L_2$ parts in the elastic net are assigned different scaling factors, we share the scaling parameter $\lambda$ (called the \textit{regularization constant}).  This choice is motivated by a need to reduce the complexity of the model selection task.  
Over a set $\calD_M$ consisting of $M$ data samples the full loss function is expressed as,
\begin{equation}
\label{eqn:full_loss}
\calL(\btheta; \lambda, \calD_M) = \frac{1}{M} \sum_{i=1}^{M} \|y_i - \hat{f}(\bxi_i, \btheta)\|^2 + \lambda  \sum_{i=1}^{L+1}\Big( \| \W^{(i)} \|_{2}^{2} + \|\W^{(i)}\|_1 \Big).
\end{equation}
From the point of view of constrained optimization, normed weight penalties limit model complexity by shrinking the feasible region of parameters, $\btheta$, to their corresponding norm balls. 
There is also a Bayesian justification for the use of normed penalties on the weights. $\btheta^*$ obtained by the minimization of the regularized loss function corresponds to the maximum a posteriori (MAP) estimate of $\btheta$, with the prior given by the chosen penalty. The $L_1$ and $L_2$ penalties correspond to a Laplace and Gaussian priors on the weights while the elastic net represents a compromise between the two. In unnormalized form, the elastic net regularizer with equi-scaling of the $L_1$ and $L_2$ parts, correspond to the following prior on the weights:
\begin{equation}
\label{eqn:prior}
p(\W) \propto \exp(\lambda \|\W\|_{2}^{2} + \lambda \|\W\|_{1}). 
\end{equation}

\subsection{Gradient computation and optimization}
\label{sec:grad_opt}
A DNN $\hat{f}(\bxi;\btheta)$ is a highly complicated function of the network parameters $\btheta$ because of the fact that it involves multiple layers of compositions of simpler functions. To perform gradient descent optimization one needs access to the gradients of the objective function. For training DNNs, this is achieved by utilizing the celebrated \textit{backpropagation} algorithm \cite{chauvin1995backpropagation}. In essence, the backpropagation algorithm is a recursive application of the standard chain rule. Unlike numerical differentiation schemes such as finite differences, backpropagation is exact. 

Training a DNN reduces to a stochastic optimization problem with the objective function being the loss function described in Eq. (\ref{eqn:full_loss}). The most common way of solving this problem is via the stochastic gradient descent (SGD) \cite{bottou2010large} algorithm. As the name suggests, SGD is the stochastic analogue of deterministic gradient descent. The SGD algorithm produces a converging sequence of updates of the optimization variables, by making appropriately sized steps in the direction of the negative gradient of the objective function. The key idea of the SGD method, is that it approximates the negative gradient of the objective function by averaging a finite set of objective function gradient samples. This is done by independently sampling a small subset of examples, $\calD_M$, from the full training dataset, $\calD$. The update scheme of the SGD method is:
\begin{equation}
\label{eqn:sgd_update}
\btheta_{k+1} \leftarrow \btheta_{k} + \alpha_k \nabla_{\btheta} \calL(\btheta; \lambda, \calD_M).
\end{equation}
 Note that the sampling of $\calD_M$ is performed at every iteration of SGD. 
While the SGD algorithm is simple to implement, it is not guaranteed to perform well for complex high dimensional objective functions (as is typical for Eq. \ref{eqn:train_dnn}). While there are multiple variants of the SGD method that have demonstrated improvements over vanilla SGD, in this work, we solve Eq. (\ref{eqn:train_dnn}) with the Adaptive Moments (ADAM) optimization algorithm \cite{kingma2014adam}. 
The ADAM update scheme is as follows:
\begin{eqnarray}
\mathbf{M}_k \leftarrow \beta_1 \mathbf{M}_{k-1} + (1-\beta_1) \mathbf{G}_k, \\
\mathbf{V}_k \leftarrow \beta_2 \mathbf{V}_{k-1} + (1 - \beta_2) \mathbf{G}_{k}^{2}, \\
\widehat{\mathbf{M}}_k \leftarrow \frac{\mathbf{M}_k}{1-\beta_{1}^{k}}, \\
\widehat{\mathbf{V}}_k \leftarrow \frac{\mathbf{V}_k}{1-\beta_{2}^{k}}, \\
\bm{\btheta}_{k+1} \leftarrow \bm{\btheta}_{k} + \alpha_k \frac{\widehat{\mathbf{M}}_k}{\sqrt{\widehat{\mathbf{V}}_k} + \eta},
\label{eqn:adam_update}
\end{eqnarray}
where, $\mathbf{G}_k = \nabla_{\btheta} \calL(\btheta; \lambda, \calD_M)$ is the estimate of the objective function gradient at iteration $k$ and $\mathbf{M}_k$ and $\mathbf{V}_k$ are exponential moving average estimates of gradients and squared gradients respectively. $\mathbf{M}_0$ and $\mathbf{V}_0$ are set to 0 and the bias introduced by this initialization is corrected by computing $\widehat{\mathbf{M}}_k$ and $\widehat{\mathbf{V}}_k$. 
$\eta$ is a suitably small number introduced to prevent 0 denominator. $\beta_1$ and $\beta_2$ are averaging parameters which can be tuned. 
In practice, default values of $\beta_1=0.9$, $\beta_2=0.999$, as suggested by \cite{kingma2014adam} work well and we do not fiddle with these quantities. 

\subsection{Selecting network structure}
\label{sec:net_struct}
Although various authors in the  literature offer rules of thumb for selecting the number and size of DNN layers (such as those suggested in \cite{bengio2012practical}), rigorous rules for the selection of these quantities do not exist.
One typically resorts to extensive experimentation to arrive at a suitable network configuration. In the most naive case, the number and size of the hidden layers are hyperparameters selected using cross-validation. In this work, we are interested in learning a surrogate of the form described in Eq. (\ref{eqn:dnn_as}). 
The function $h$ accepts an input in a vector space of dimensions $D$ and projects it to a vector space of dimension $d$, where $d << D$ ($d$ is to be determined through our methodology). 
We parameterize this section of the DNN such that the widths of it's hidden layers decays exponentially. Specifically, the number of hidden units in the $k^{th}$ hidden layer in this section is given by:
\begin{equation}
\label{eqn:net_structure}
d_k = \ceil{D \exp(\rho k)},
\end{equation}
where, $\ceil{a}$ represents the ceiling (closest greater integer) of the number $a$. The parameter $\rho$ is uniquely determined.  
\begin{wrapfigure}{r}{0.45\textwidth}
\includegraphics[width=0.45\textwidth]{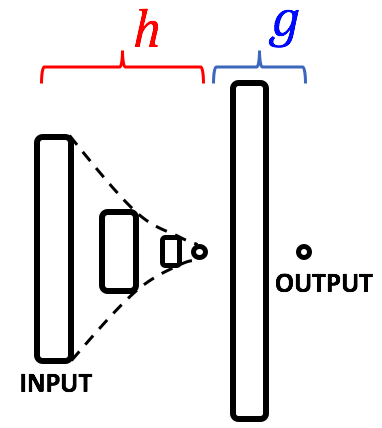}
\caption{Visualization of the parameterized network structure with $L=3$ and $d=1$.}
\label{fig:net_vis}
\end{wrapfigure}
The link function $g$ is formulated as a single layer MLP. The hidden layer in $g$ is taken to have a width of $300d$. One could set this width to be anywhere between $100 - 500$ times the size of the encoding $d$. The idea is that the subnetwork representing the link function $g$ ought to have a sufficient number of hidden units to capture arbitrary nonlinearities. A visual representation of the DNN surrogate is shown in Fig. \ref{fig:net_vis}. Note that no activation function is used at the output of the encoding subnetwork $h$, and the output of the link function subnetwork, $g$. The task of optimizing the network structure is then reduced to a task of cross validating over two integer quantities, $L$ and $d$, a much simpler task than optimizing for the number of layers and sizes of the individual layers separately.  

\subsection{Combined global optimization and grid search for model selection}
\label{sec:full_proc}
The stochastic optimization task stated in Eq. (\ref{eqn:train_dnn}) is characterized by hyperparameters, weight decay $\lambda$, and the integer quantities $L$ and $h$, which fully parameterize the structure of the network. We refer to structure parameters collectively, as, $\mathcal{S} = (L, h)$. Training a DNN involves, in addition to optimizing for $\btheta$, selection of appropriate values of hyperparameters. The naive approach to do this is to perform an intuition guided manual search. In this work, the task of model selection reduces to selecting 3 quantities - the discrete hyperparameters, $L$ and $d$ and the continuous hyperparameter, $\lambda$. To be systematic, we adopt a combined grid search and stochastic global optimization procedure. Specifically, we define a grid of values for $L$ and $d$. Over each grid location of the structure parameters, we perform a Bayesian global optimization  (BGO) \cite{brochu2010tutorial,pandita2016extending} for $\lambda$. 

We split the dataset $\calD$ into 3 parts - a training set, $\calD_{\mathrm{train}}$, a validation set, $\calD_{\mathrm{val}}$ and a test set, $\calD_{\mathrm{test}}$. We define a grid, $\mathcal{G}$, of $L$ and $h$ values and seek to assign a score to each location on the grid. The optimal choice of the structure parameters, $\calS$ would then be the grid location which minimizes the validation error:
\begin{equation}
\label{eqn:score}
\calR(\calS; \lambda) = \frac{1}{M_{\mathrm{val}}}\sum_{i=1}^{M_{\mathrm{val}}}(y_{\mathrm{val}, i} - \hat{f}_{\calS}(\bxi_{\mathrm{val}, i};\btheta_{\calS}^{*}(\lambda)))^2,
\end{equation}
where, $(\bxi_{\mathrm{val}, i}, y_{\mathrm{val}, i}) \in \calD_{\mathrm{val}}$, $M_{\mathrm{val}}$ is the size of the validation set, and $\hat{f}_{\mathcal{S}}$ is a DNN characterized by structure parameter, $\mathcal{S}$. $\btheta_{\mathcal{S}}^{*}(\lambda)$ is an estimate of the network parameters, $\btheta$ obtained by minimizing the loss function in Eq. (\ref{eqn:full_loss}), with the regularization constant set to $\lambda$ and network structure parameter, $\calS$. 

The optimal choice of regularization constant $\lambda$, corresponding to structure parameter, $\mathcal{S}$, is:
\begin{equation}
\label{eqn:const_Struct_opt_lambda}
\lambda^{*}_{\mathcal{S}} = \underset{\lambda}{\arg\min} \ \E [\calR (\mathcal{S}, \lambda)]. 
\end{equation}
Eq. (\ref{eqn:const_Struct_opt_lambda}) is a stochastic global optimization problem characterized by a noisy objective function.
BGO sequentially seeks out the global optimum of the objective function, $\calR$, by iteratively updating a Gaussian process (GP) surrogate response surface for $\calR(\lambda;\calS)$. During each iteration of BGO, a new pair of input-output observations is generated by maximizing an information acquisition function (IAF). The most popular choice of IAF is the expected improvement (EI) function. In closed form, the EI-IAF is given by:
\begin{equation}
\label{eqn:ei}
\mathrm{EI}(\lambda) = 
\begin{cases}
(\mu(\lambda) - \calR(\lambda^{*}; \calS)) \Phi(Z) + \sigma(\lambda) \phi(\lambda)\ \text{if } \sigma(\lambda) > 0, \\
0 \ \text{if } \sigma(\lambda) \leq 0,
\end{cases}
\end{equation}
where, $\phi$ and $\Phi$ are the probability density function and the cumulative distribution function of the standard normal distribution. $Z = \frac{\mu(\lambda) - \calR(\lambda^*; \calS)}{\sigma(\lambda)}$ where, $\mu(\lambda)$ is the predictive mean of the GP surrogate at $\lambda$, and $\sigma(\lambda)^2 = \sigma_{\mathrm{GP}}(\lambda)^2 - \sigma_{\mathrm{noise}}(\lambda)^2$, where $\sigma_{\mathrm{GP}}(\lambda)^2$ is the predictive variance of the GP surrogate which captures the epistemic uncertainty induced due to the limited set of observations and $\sigma_{\mathrm{noise}}(\lambda)^2$ is GP estimate of the observational noise induced by random initializations of the DNN  weights and random splitting of the dataset into $\calD_{\mathrm{train}}$, $\calD_{\mathrm{test}}$ and $\calD_{\mathrm{val}}$. $\sigma(\lambda)^2$ is thus a filtered version of the predictive variance which is robust to observational noise. The BGO algorithm is summarized in Alg. \ref{alg:bgo_opt}. Note that the we maximize the negative of the validation error $\calR$.

\begin{algorithm}
\caption{Bayesian global optimization of validation error $\calR(\lambda, \calS)$} \label{alg:bgo_opt}
\begin{algorithmic}[1]
\Require Training data, $\calD_{\mathrm{train}}$, validation data, $\calD_{\mathrm{val}}$, structure parameter, $\calS$, number of initial observations, $n_{\mathrm{init}}$, number of BGO iterations, $maxiter$, bounding box for $\lambda$, $\mathcal{B}$.
\State Initialize empty arrays, $\Lambda_{\mathrm{BGO}}$ and $\calR_{\mathrm{BGO}}$.
\State Use Latin Hypercube sampling (LHS) \cite{iman2008latin} to generate $n_{\mathrm{init}}$ samples of $\lambda$ within the bounding box $\mathcal{B}$. Call it $\Lambda_{\mathrm{init}}$.  
\For{$\lambda_i \in \Lambda_{\mathrm{init}}$}  
\State Solve Eq. (\ref{eqn:full_loss}) with training data, $\calD_{\mathrm{train}}$ to obtain $\btheta_{\calS}(\lambda_i)$. 
\State Evaluate $\calR_i$ = -$\calR(\lambda_i;\calS)$. 
\State Append $\lambda_i$ and $\calR_i$ to $\Lambda_{\mathrm{BGO}}$ and $\calR_{\mathrm{BGO}}$ respectively. 
\EndFor
\State Fit a GP surrogate linking $\Lambda_{\mathrm{BGO}}$ and $\calR_{\mathrm{BGO}}$. 

\For{iter = 1 to $maxiter$}
\State Get next sample of $\lambda$, $\lambda_{n_{\mathrm{init}} + iter} = \underset{\lambda}{\arg\max} EI(\lambda)$. 
\State Evaluate $\calR_{n_{\mathrm{init}} + iter} = -\calR(\lambda_i;\calS)$. 
\State Append $\lambda_{n_{\mathrm{init}} + iter}$ and $\calR_{n_{\mathrm{init}} + iter}$ to  $\Lambda_{\mathrm{BGO}}$ and $\calR_{\mathrm{BGO}}$ respectively.
\State Update GP surrogate based on augmented dataset, $\Lambda_{\mathrm{BGO}}$ and $\calR_{\mathrm{BGO}}$.
\EndFor

\State Get $index = \arg\max{\calR_{\mathrm{BGO}}}$. \\
\Return $\lambda^{*}_{\calS} = \Lambda_{\mathrm{BGO}}(index)$. \Comment{Return $\lambda$ corresponding to the highest observed negative validation error.}
\end{algorithmic}
\end{algorithm}
Finally, the optimal structure parameter, $\calS^{*}$ is given by: 
\begin{equation}
\label{eqn:opt_struct_param}
\mathcal{S}^{*} =  \underset{\calS \in \mathcal{G}}{\arg\min} \ \calR(\mathcal{S}, \lambda^{*}_{\calS}).
\end{equation}
The full algorithm is summarized in Alg. \ref{alg:train_dnn}. The estimation of $\calR_{\calS}$ for each individual $\calS$ can be parallelized and the computational cost of the global optimization search for $\lambda^{*}_{\calS}$ requires $maxiter + n_{\mathrm{init}}$ times the cost of a single run of the ADAM optimizer. 

\begin{algorithm}
\caption{Full procedure for training DNN surrogate} \label{alg:train_dnn}
\begin{algorithmic}[1]
\Require Data, $\mathcal{D}=(\X, \y)$, grid of $L$ and $h$ values, $\mathcal{G}$, parameters for Alg. \ref{alg:bgo_opt}, $n_{\mathrm{iter}}$, $maxiter$ and $\mathcal{B}$.
\State Split $\calD$ into 3 parts - $\calD_{\mathrm{train}}, \calD_{\mathrm{test}}$ and $\calD_{\mathrm{val}}$. 
\For{$\mathcal{S}  \in \mathcal{G}$}
\State Set $\lambda^{*}_{\mathcal{S}} = \underset{\lambda}{\arg\min} \ \mathbb{E}[\calR (\calS, \lambda)]$. \Comment{Using Alg. \ref{alg:bgo_opt}}.
\State Set $\btheta^{*}_{\calS, \lambda} \leftarrow \btheta^{*}_{\calS}(\lambda_{\mathcal{S}}^{*})$. 
\State Set $\calR_{\calS} \leftarrow \calR(\mathcal{S}, \lambda_{\mathcal{S}}^{*};\btheta_{\mathcal{S}, \lambda}^{*})$.
\EndFor
\State Set $\calS^{*} \leftarrow \underset{\mathcal{S} \in \mathcal{G}}{\arg\min} \  \calR_{\calS}$. \Comment{Get the structure parameter that minimizes the observed validation error.}
\State \Return $\calS^{*}, \lambda_{\calS^*}^{*}, \btheta_{\calS^{*}, \lambda_{\calS^*}^{*}}^{*}$.  \Comment{Final DNN surrogate.}
\end{algorithmic}
\end{algorithm}

%% file: examples.tex
\section{Numerical Example - Stochastic Elliptic Partial Differential Equation}
\label{sec:examples}


We consider the following benchmark elliptic PDE on the 2-d unit square domain:
\begin{equation}
\label{eqn:elliptic_pde}
-\nabla (a(\x) \nabla u(\x)) = 0, \ \forall \x \in [0, 1]^2,
\end{equation}
with boundary conditions:
\begin{eqnarray}
\label{eqn:pde_bc}
u = 0,\  \forall x = 1, \\ 
u = 1,\ \forall x = 0, \\
\frac{\partial u}{\partial n} = 0,\ \forall y = 0 \text{ and } y = 1,
\end{eqnarray}
where, $\x = (x, y)$ are the physical coordinates in 2-d Euclidean space. 

Eq. (\ref{eqn:elliptic_pde}) is a model for 2-d steady-state diffusion processes. The quantity $a(\x)$ is a spatially varying diffusion coefficient. 
The physical significance of the equation and all terms in it are derived from context. For instance, Eq. (\ref{eqn:elliptic_pde}) could be an idealized model for single-phase groundwater flow in an aquifer \cite{li2016inverse}, where $a$ represents the transmissivity coefficient and the solution variable, $u$ is the pressure. 

It is often the case that the $a$ is unknown throughout the PDE domain. The uncertainty in the diffusion coefficient is formalized by modeling $a$ as a log normal random field, i.e.,  i.e., 
\begin{equation}
\label{eqn:random_field_gp}
\log a(\x) \sim \mathrm{GP}(a(\x)|m(\x), k(\x, \x')), 
\end{equation}
where, $m(\x)$ and $k(\x, \x')$ are the mean and covariance functions, respectively, of the Gaussian random field which models the logarithm of the diffusion coefficient $a(\x)$. The mean function models beliefs about the generic trends of the diffusion field as a function of spatial location. For the sake of simplicity, we set $m(\x) = 0$ in this example. The covariance function $k$ models beliefs about the regularity of the diffusion field and the and the lengthscales in which it varies. A popular choice for $k$ is the exponential kernel:

\begin{equation}
\label{eqn:exp_cov_func}
k(\x, \x') = \exp\bigg(- \large\sum_{i=1}^{2} \frac{|x_i - x'_{i}|}{\ell_i}\bigg),
\end{equation}
where $\ell_i$ represents the correlation length along the $i^{th}$ spatial direction.
The correlation lengths are typically assigned a fixed value. One then proceeds to use the truncated KLE to produce a reduced representation of the infinite-dimensional random field. The coefficients of the KLE are i.i.d. standard normal, and realizations of the diffusion field, $a$, can be generated easily, i.e.,  by sampling the KLE coefficients. For each realization of $a$, the corresponding solution of Eq. (\ref{eqn:elliptic_pde}) is obtained. Any relevant quantity of interest, $q = \mathcal{Q}[u]$, is computed.  Finally, one learns a surrogate response surface that maps the coefficients of the truncated KL expansion,  using suitable learning algorithms such as GP regression. 

We broaden the scope of the problem by removing the restrictions on the lengthscales, $\ell_i$. The goal, in this example, is to construct a surrogate, which can accurately predict the solution, $u$, of the PDE, regardless of the lengthscales of the realization of $a$. Our approach, then, is to construct a surrogate which directly maps the discretized random field, to the numerical solution of the PDE.  

\subsection{Forward model}
\label{sec:ex_forward_model}
We solve the PDE using the finite volume method (FVM). The solver is implemented in the \texttt{Python} library \texttt{FiPy} \cite{guyer2009fipy}.
The unit square domain is discretized into $32 \times 32$ finite volume cells. The input to the solver, $\hat{\ax} \in \R^{32 \times 32}$ is the discretized version of a sample of the random diffusion $a$. The output of the solver, $\hat{\ux} \in \R^{32 \times 32}$, is the numerical solution of the PDE corresponding to the realization $\hat{\ax}$ of the diffusion field. 

The model inputs, $\bxi = (\mathrm{vec}(\hat{\ax}), \x) \in \R^{1026}$ are spatial coordinates appended to a flattened version of the discrete random field realization and the model outputs are the PDE solution at the FV cell centers, $u(\x, \hat{\ax})$. Our goal is to learn a surrogate response, $\hat{f}:\R^{1024} \times [0, 1]^2 \rightarrow \R$, which maps the snapshot of the diffusion field and a particular coordinate in the unit square to the solution of the PDE at that location. 

\subsection{Data Generation}
\label{sec:ex_data_gen}
Intuitively, we would like to sample more realizations, of $a$, that have smaller lengthscales because one would observe the most variability in the solution corresponding to low lengthscale diffusion fields. Thus, instead of sampling lengthscale pairs uniformly from the unit square, we bias our sampling procedure to pick lengthscales that are smaller. Alg. \ref{alg:gen_ls} describes the procedure to select lengthscales to train the DNN surrogate. Note that a lower bound on $\ell_i$ is set by constraining the lengthscale to be larger the FV cell size, $h(= \frac{1}{32})$. 

\begin{algorithm}
\caption{Sampling of lengthscale pairs.}\label{alg:gen_ls}
\begin{algorithmic}[1]
\Require Number of lengthscale pairs, $n$, lower bound on lengthscale, $h$. 
\State Initialize $n$-dimensional empty array $\mathbf{L}$.
\State Initialize counter, $c = 1$.
\While{$c \leq 60$}
\State Sample $u = (u_1, u_2, u_3) \sim \mathcal{U}([0, 1]^3)$. \Comment{$\mathcal{U}(\mathcal{A})$ is the uniform distribution over the set $\mathcal{A}$.}
\If{$\exp(-u_1 - u_2) < u_3$}
\State Set $\ell_c = (\ell_{x, c}, \ell_{y, c}) = (h+u_1(1-h), h+u_2(1-h))$. \Comment{Scale the sampled lengthscales to the range $[h, 1]$.}
\State Set $\mathbf{L}_c \leftarrow  \ell_c$.
\State Increment counter $c \leftarrow c + 1$.
\EndIf
\EndWhile
\Return $\mathbf{L}$. 
\end{algorithmic}
\end{algorithm}
We generate $n$ different lengthscale pairs following the procedure in Alg. \ref{alg:gen_ls} to obtain a design of lengthscale pairs $\mathbf{L}$. 

\begin{figure}[h]
\includegraphics[width=\textwidth]{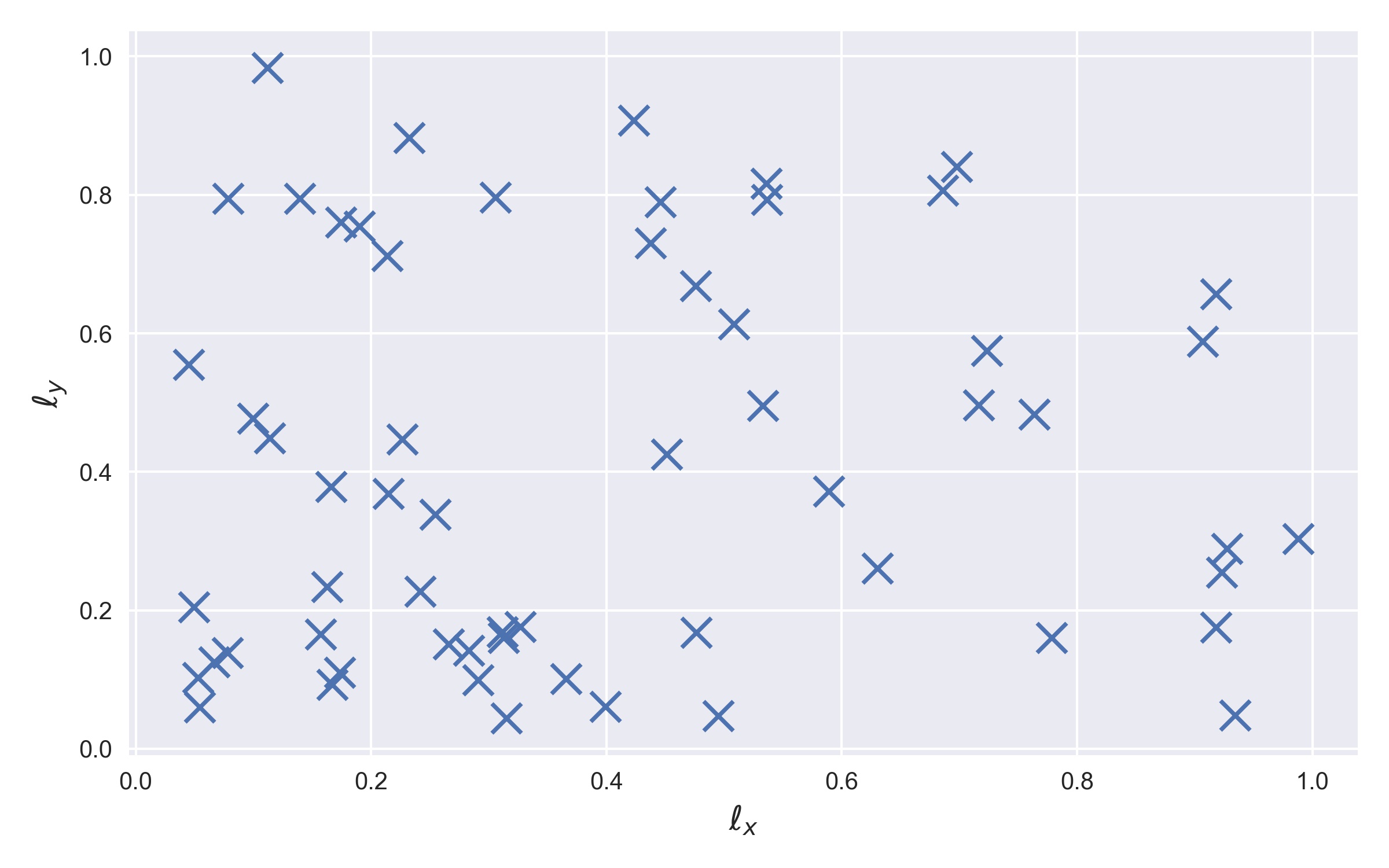}
\caption{Visual representation of LHS design of lengthscale pairs. Each 'x' represents a sampled pair of lengthscales.}
\label{fig:sampled_lx}
\end{figure}
\begin{figure}[h]
\includegraphics[width=\textwidth]{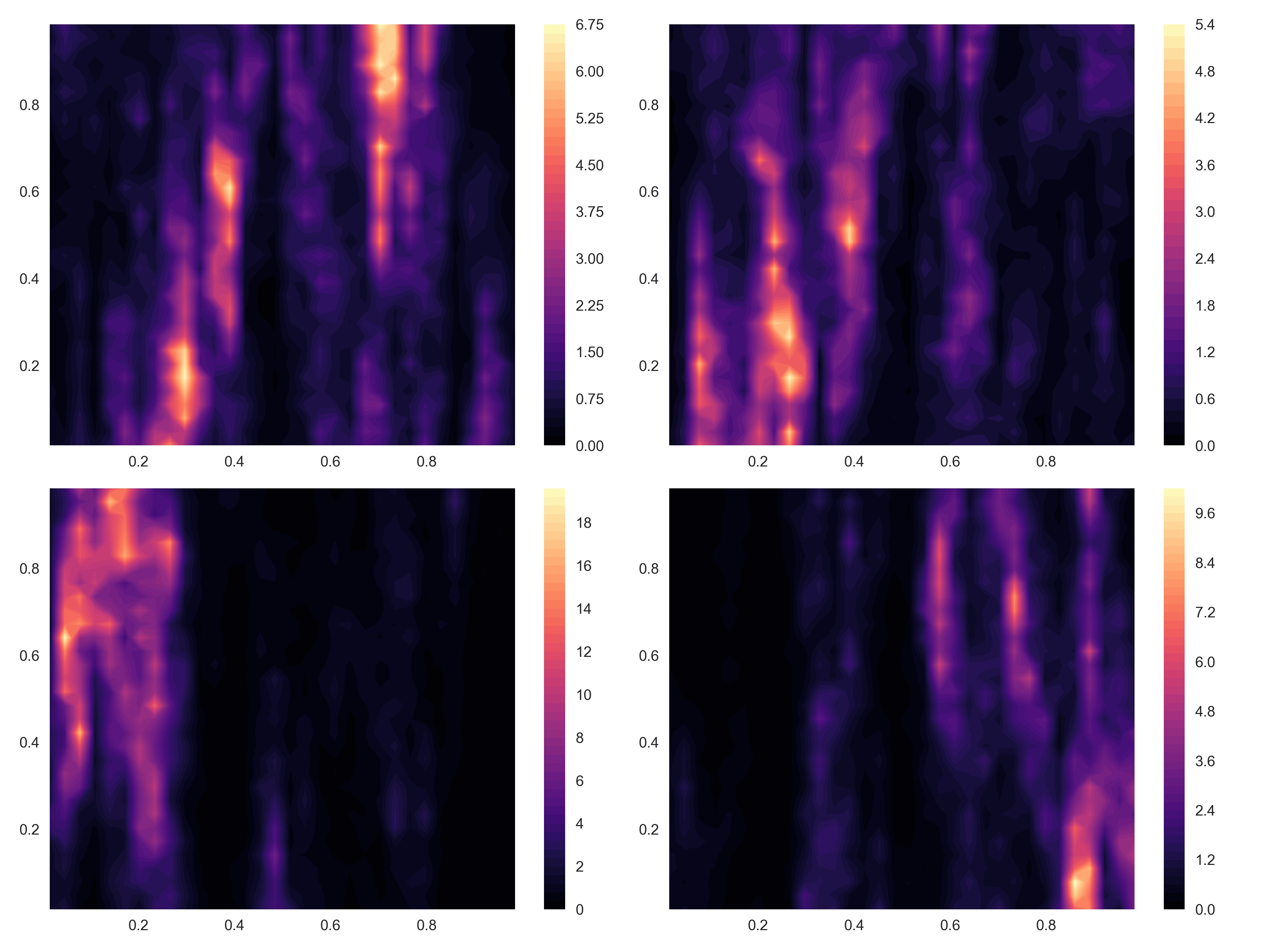}
\caption{Samples of the random field $a(\x)$ with lengthscales $\ell_x = 0.446$ and $\ell_y = 0.789$ along the $x$ and $y$ directions.}
\label{fig:sample_diffusion_1}
\end{figure}
\begin{figure}[h]
\includegraphics[width=\textwidth]{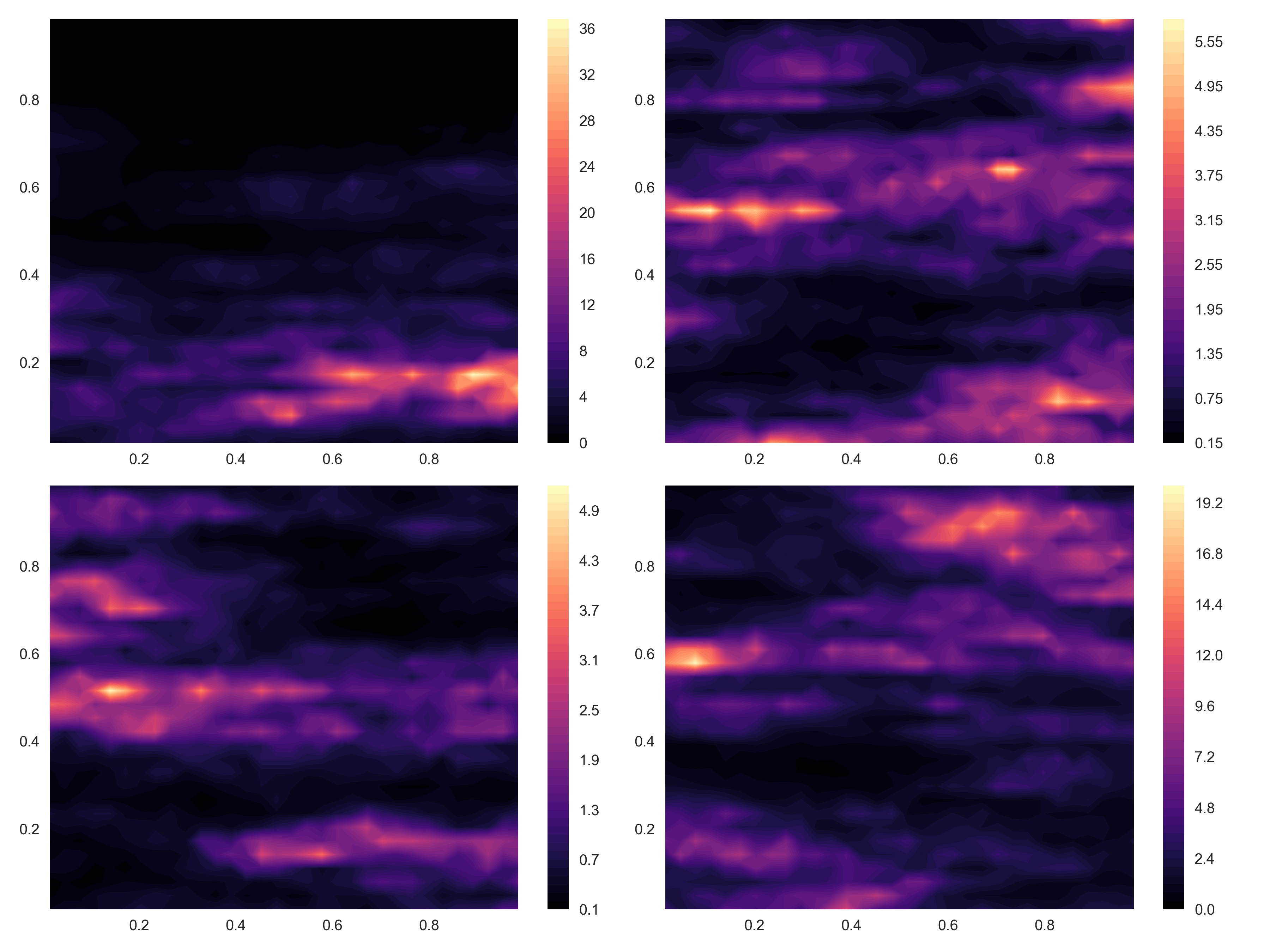}
\caption{Samples of the random field $a(\x)$ with lengthscales $l_x = 0.291$ and $l_y = 0.099$ along the $x$ and $y$ directions.}
\label{fig:sample_diffusion_2}
\end{figure}

For each lengthscale pair, $(\ell_x, \ell_y) \in \mathbf{L}$, we solve the forward model $N$ times by generating $N$ realizations of the diffusion coefficient. In this example we set $n = 60$ and $N=100$. A visual representation of $\mathbf{L}$ is shown in Fig. \ref{fig:sampled_lx}. The full data generation procedure is summarized in Alg. \ref{alg:data_gen}. Samples of the diffusion coefficient drawn from two different pairs of lengthscales are shown in Fig. \ref{fig:sample_diffusion_1} and \ref{fig:sample_diffusion_2}.

\begin{algorithm}
\caption{Data generation.}\label{alg:data_gen}
\begin{algorithmic}[1]
\Require Number of unique pairs of lengthscales, $n$, and number of samples per lengthscale pair, $N$. 
\State Use Alg. \ref{alg:gen_ls} to generate $n$ pairs of lengthscales.  
\For{$(\ell_x, \ell_y) \in \mathbf{L}$} 
\State Generate $N$ samples of $a$ with lengthscales $\ell_x$ and $\ell_y$ along $x$ and $y$ directions respectively.
\State Run FV PDE solver to generate solutions corresponding to each sample of $a$.
\EndFor
\end{algorithmic}
\end{algorithm}


\subsection{Numerical settings}
\label{sec:num_settings}
\subsubsection{Dataset split}
\label{sec:data_split}
We generated our dataset, $\calD$, of $n \times N = 60 \times 100 = 6000$ pairs of $\hat{a}$ and $\hat{\ux}$, based on the procedure outlined in Alg. \ref{alg:data_gen}. $\calD$ is randomly shuffled and split into 3 parts - A set of 2000 training examples, $\calD_{\mathrm{train}}$, a set of 2000 validation examples, $\calD_{\mathrm{val}}$ and a set of 2000 test examples, $\calD_{\mathrm{test}}$. For the purpose of constructing the surrogate, we work with the logarithm of both $\hat{\ax}$ and $\hat{\ux}$ during training. Furthermore, $\hat{\ux}$ in the training set is standardized along each dimension. Necessary inversions of the transformations are performed during test time. 

\subsection{Model selection settings}
\label{sec:model_selection_settings}
For selection of $\lambda^{*}_{\calS}$ using Alg.  \ref{alg:bgo_opt}, we set the number of initial design points, $n_{\mathrm{init}} = 5$. The number of BGO iterations, $maxiter = 10$ and the bounding box, $\mathcal{B} = [10^{-10}, 10^{-3}]$. The grid of structure parameters is set to be $\calG = \{ 3, 4, 5, 6, 7, 8,  9\} \times \{1, 2, 3, 4 \}$. 

\subsubsection{Network optimizer settings}
\label{sec:opt_set}
We set the ADAM optimization learning rate $\alpha$ to be $1 \times 10^{-3}$. The optimizer is run for $45000$ iterations and $\alpha$ is decreased by a factor of $0.1$ every $15000$ iterations. The batch size, $M$, is set to be $50$. Default values of tunable parameters of the ADAM optimizer are used, as recommended in \cite{kingma2014adam}. These settings are, by no means, universal. Refer to \cite{bengio2012practical} for some practical guidelines on DNN hyperparameter selection. 

We use the \texttt{Python} library \texttt{tensorflow} to write scripts for training our DNN surrogates. For the purpose of reproducibility, the \texttt{NumPy} pseudo-random number generator seed is fixed. 
The code to replicate all the results in this paper will be made available at \url{https://github.com/rohitkt10/deep-uq-paper} upon publication of this manuscript.

\subsection{Results}
\label{sec:model_sel_res}
\begin{figure}[h]
\subfigure[]{
\includegraphics[width=0.48\textwidth]{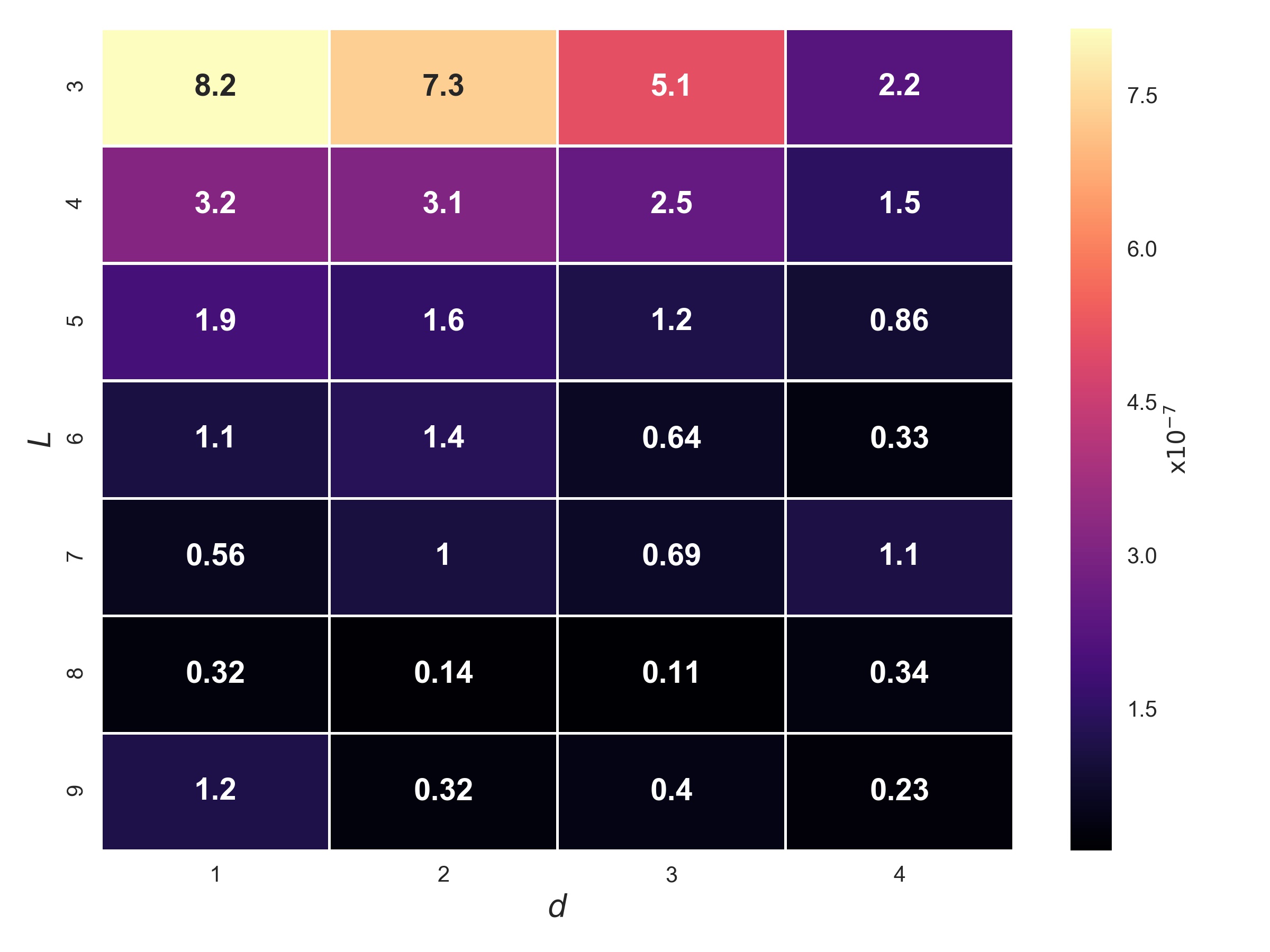}
\label{fig:N_2000_lambda_heatmap}
}
\subfigure[]{
\includegraphics[width=0.48\textwidth]{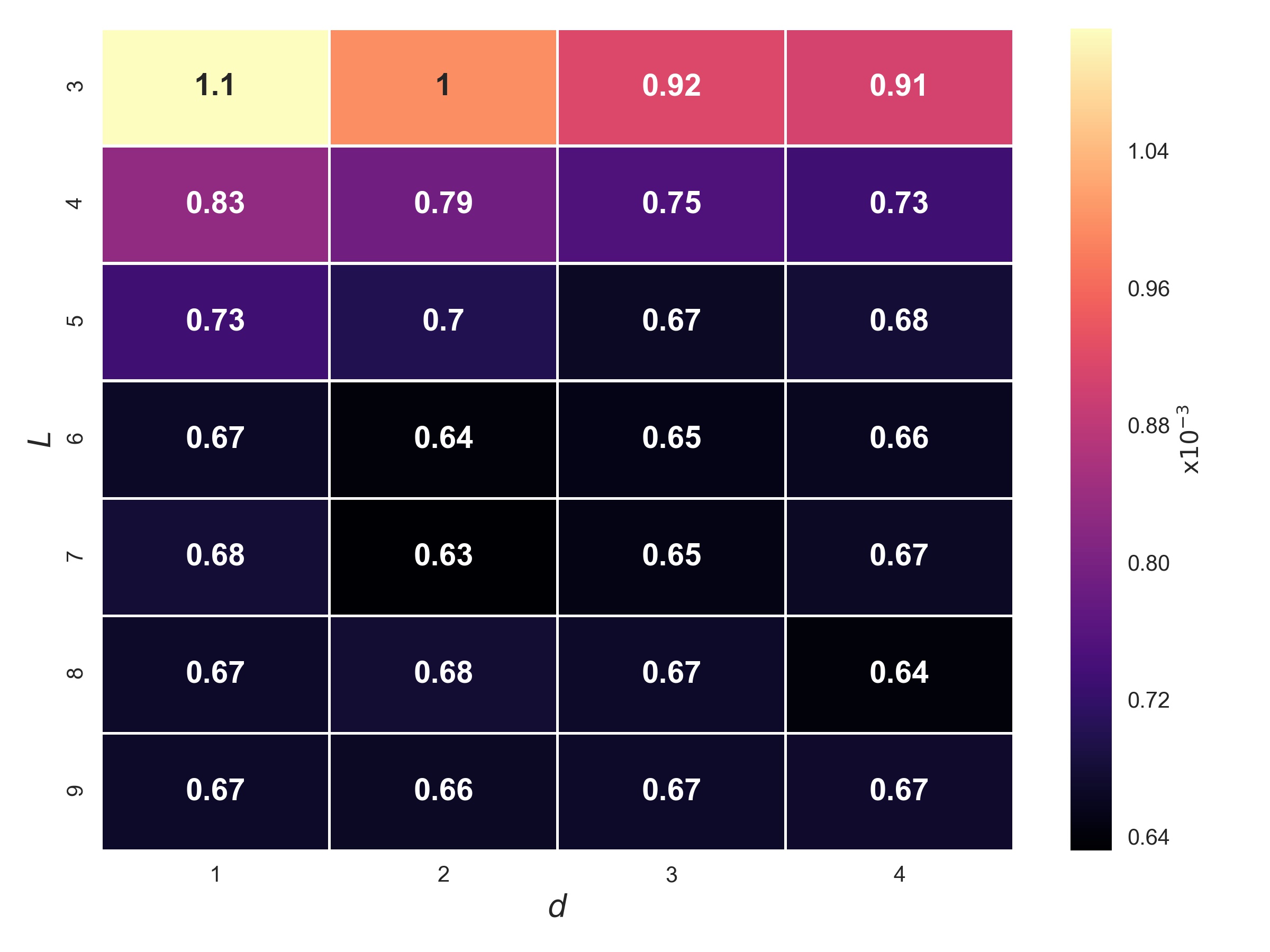}
\label{fig:N_2000_r_heatmap}
}
\caption{ \ref{fig:N_2000_lambda_heatmap} - Heatmap of $\lambda_{\calS}^{*}$ over the grid $\calG$. \ref{fig:N_2000_r_heatmap} - Heatmap of $\calR_{\calS}$ over the grid $\calG$.}
\label{fig:model_sel_heatmaps_n_2000}
\end{figure}

\begin{figure}[h]
\includegraphics[width=\textwidth]{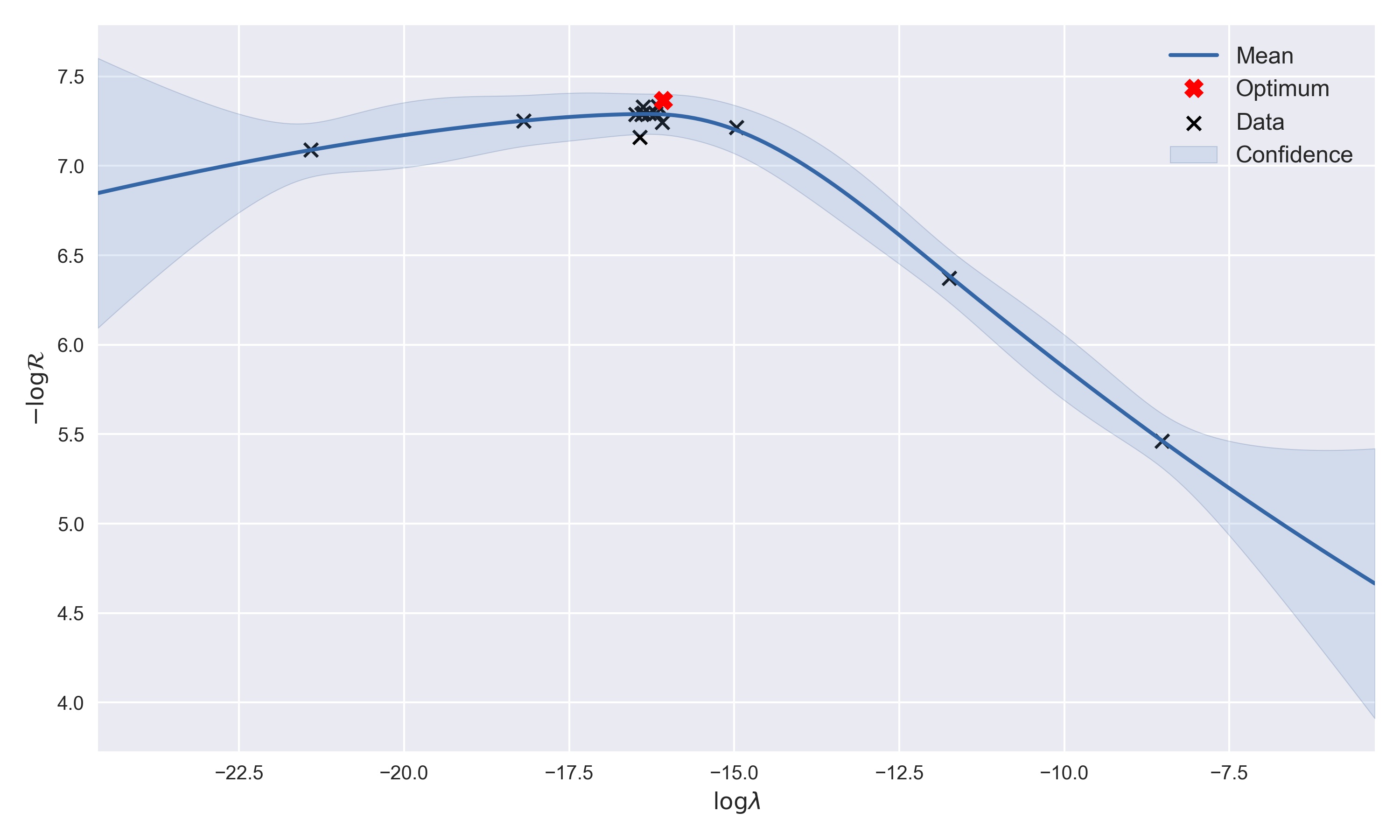}
\caption{Gaussian process surrogate generated during BGO. We maximize the negative of the validation error $\calR$ as a function of the logarithm of the regularization parameter, $\lambda$.}
\label{fig:bgo_gp}
\end{figure}

Fig. \ref{fig:model_sel_heatmaps_n_2000} shows a heatmap of $\lambda$ and optimal validation error $\calR_{\calS}$ over the grid of the structure parameters. We observe that for the chosen grid, the optimal structure parameter is found to be $\calS^{*} = (7, 2)$ and $\lambda_{(7, 2)}^{*} = 1.043 \times 10^{-7}$. Fig. \ref{fig:bgo_gp} shows GP surrogate response for $-\log \calR$ as a function of $\log \lambda$, for $\calS = (7, 2)$. Observe, from Fig. \ref{fig:bgo_gp} that there is a dense clustering of the 'x' markers around the optimum, indicating the convergence of the sequential optimization process.

\begin{figure}[h]
\includegraphics[width=\textwidth]{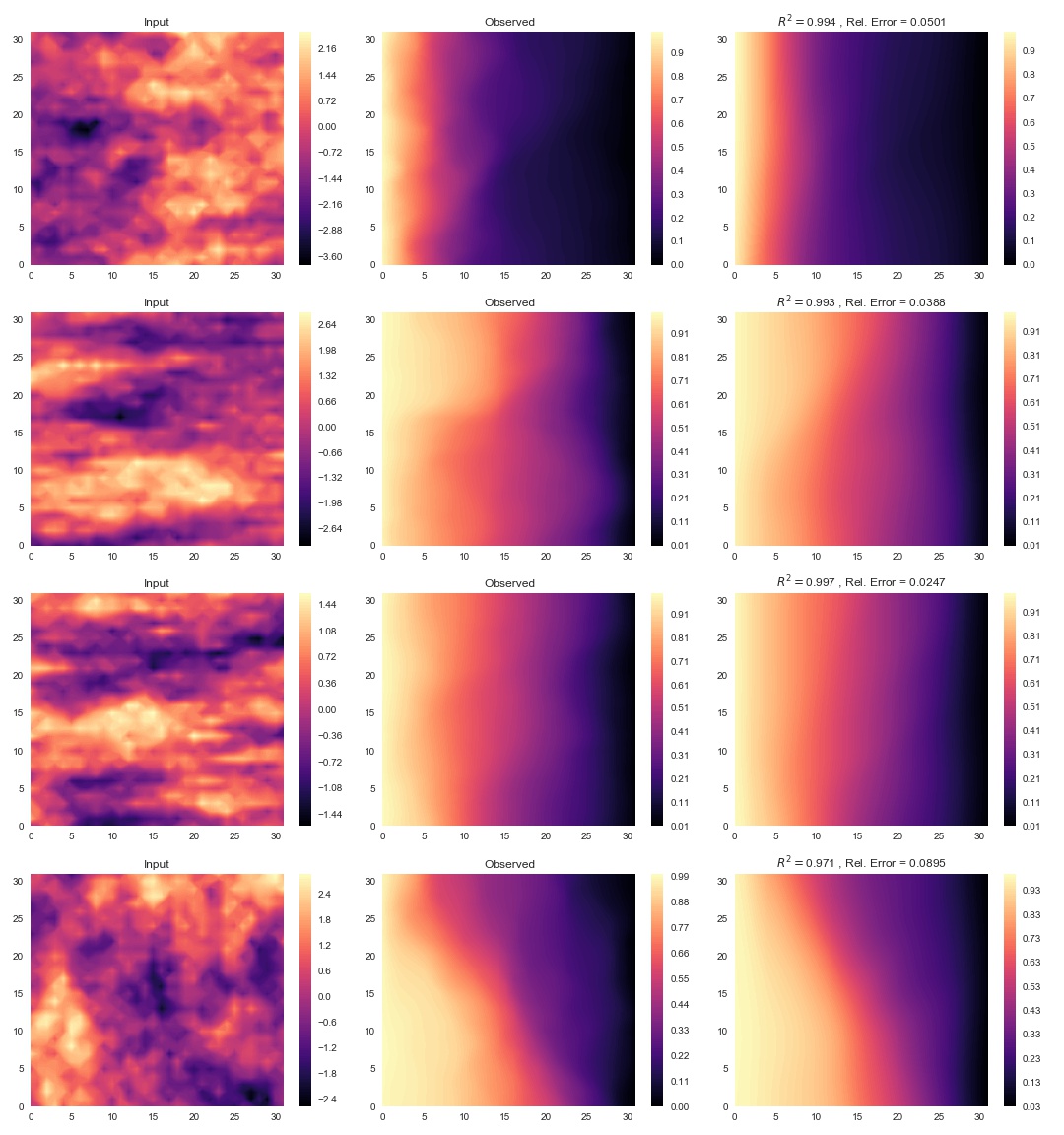}
\caption{Comparisons of DNN prediction of the PDE solution to that correct solution for 4 randomly chosen test examples. The left column shows the input diffusion field, the middle column shows the FV solution of the PDE and the right column shows the solution of the PDE predicted by the DNN.}
\label{fig:pred_vs_obs}
\end{figure}

\begin{figure}[h]
\subfigure[]{
\includegraphics[width=0.48\textwidth]{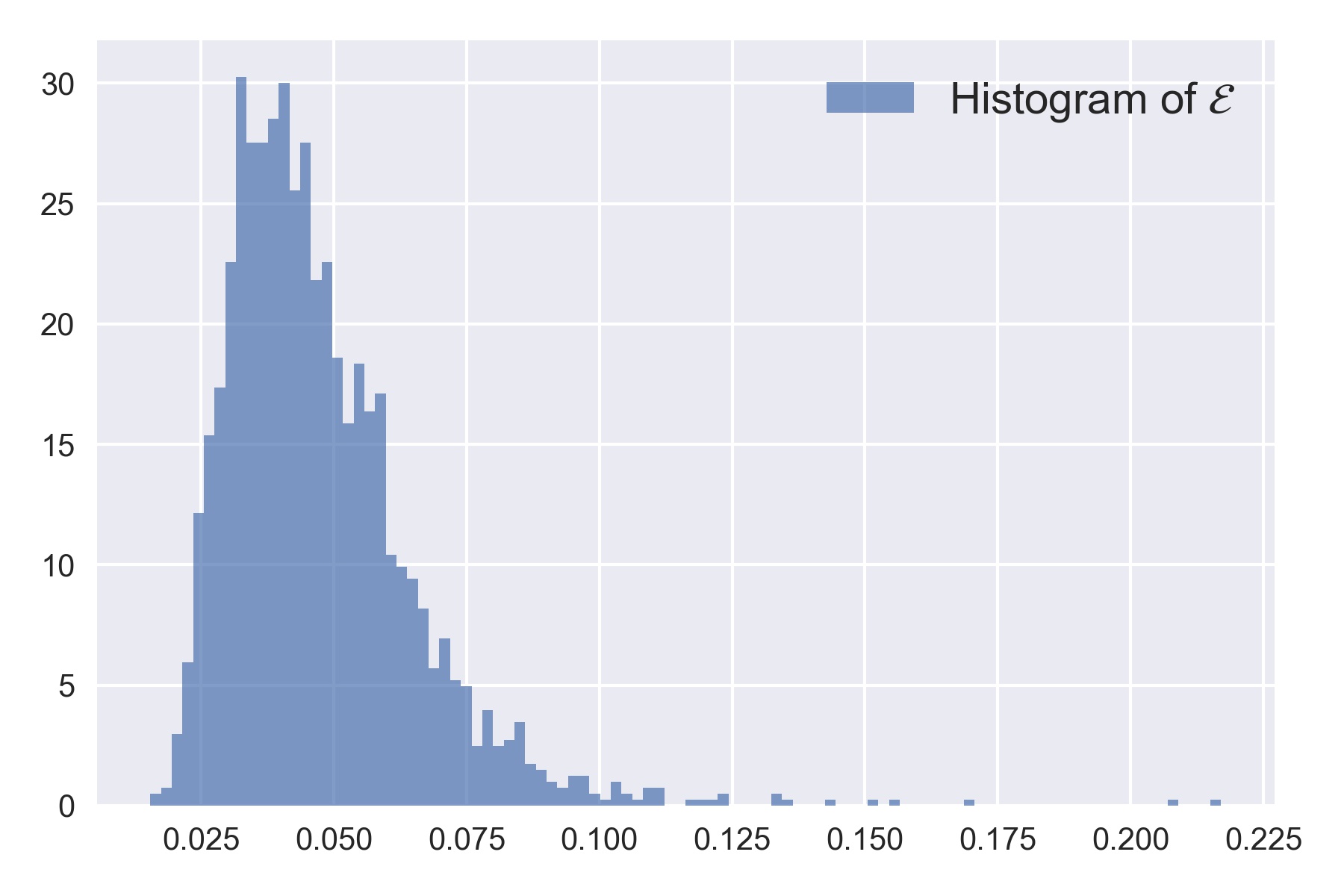}
\label{fig:rel_errors_hist}
}
\subfigure[]{
\includegraphics[width=0.48\textwidth]{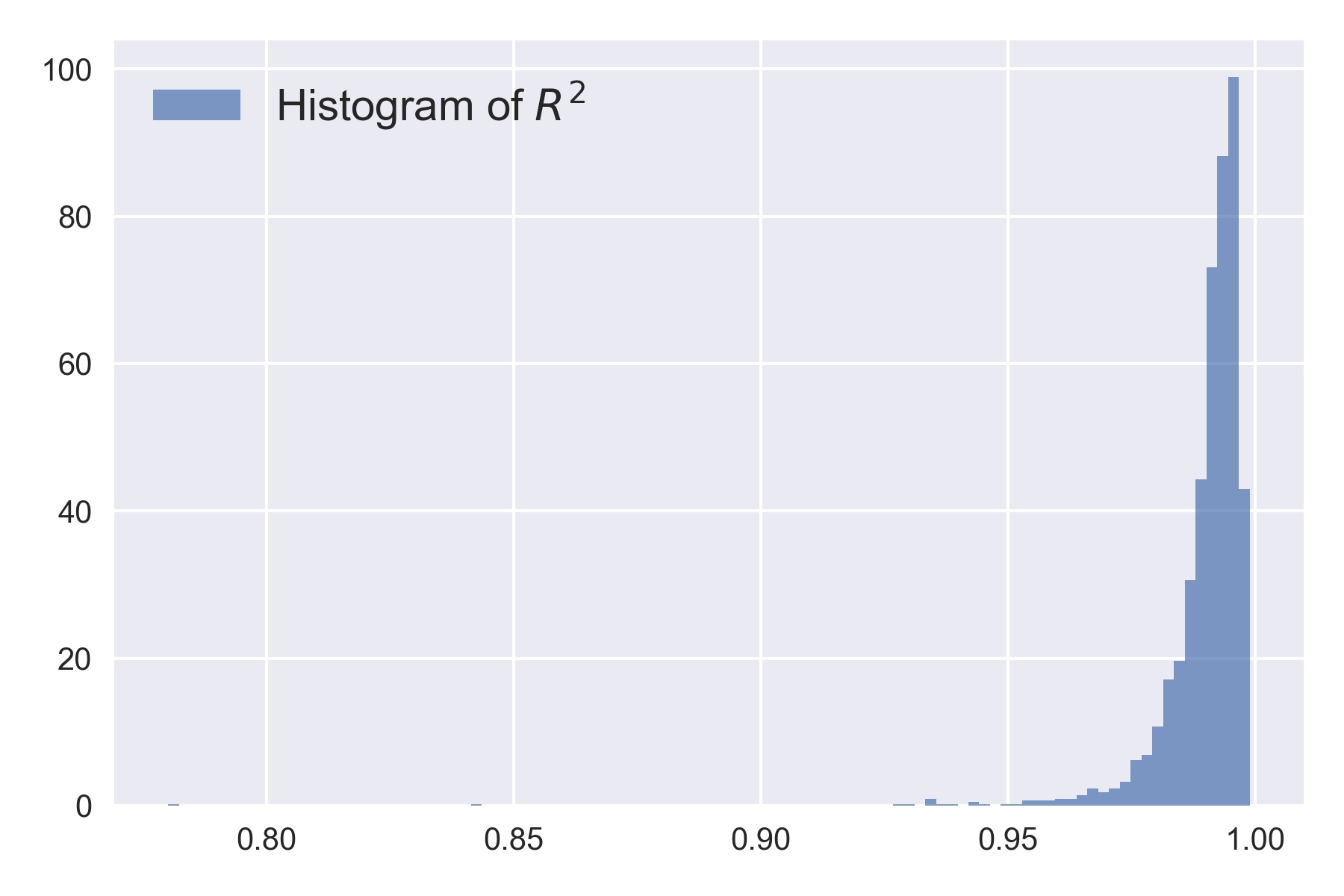}
\label{fig:r2_scores_hist}
}
\caption{\ref{fig:rel_errors_hist} - Histogram of relative errors, $\mathcal{E}$, for all examples in the test data set. \ref{fig:r2_scores_hist} - Histogram of the $R^2$ scores for all examples in the test data set.}
\label{fig:errors_hist}
\end{figure}

The quality of the DNN predictions are evaluated based on the following relative error metric:
\begin{equation}
\label{eqn:rel_error_norm}
\mathcal{E}(\hat{\ax}) = \frac{\|\hat{\ux}_{\mathrm{DNN}} - \hat{\ux}_{\mathrm{FV}} \|_{\mathrm{F}}}{\| \hat{\ux}_{\mathrm{FV}} \|_{\mathrm{F}}}, 
\end{equation}
where, $\|\cdot \|_{\mathrm{F}}$ is the Frobenius norm. $\hat{\ux}_{\mathrm{FV}}$ and $\hat{\ux}_{\mathrm{DNN}}$ are the FVM PDE solution and the DNN prediction of the PDE solution corresponding to the realization $\hat{\ax}$ of the diffusion field. We also check the coefficient of determination, (also known as the $R^2$ score), which is defined as:
\begin{equation}
\label{eqn:r2_score}
R^2 = 1 - \frac{\sum_{k=1}^{1024}(\hat{\ux}_{\mathrm{FV}, k} - \hat{\ux}_{\mathrm{DNN}, k})^2}{\sum_{k=1}^{1024}(\hat{\ux}_{\mathrm{FV}} - \bar{\ux}_{\mathrm{FV}})^2},
\end{equation}
where, $k$ indexes all the FV cell centers, $\hat{\ux}_{\mathrm{FV}, k}$ and $\hat{\ux}_{\mathrm{DNN}, k}$ are the FV solution and DNN predicted solution at the $k^{th}$ cell center respectively, and $\bar{\ux}_{\mathrm{FV}}$ is the mean of $\hat{\ux}_{\mathrm{FV}}$. 
Fig. \ref{fig:pred_vs_obs} shows a comparison of the DNN predicted PDE solution solution corresponding to a few randomly chosen realizations of the diffusion field from $\calD_{\mathrm{test}}$. 
We observe that the relative error as reported on the headers of the predicted fields in Fig. \ref{fig:pred_vs_obs} are less than $ 0.1$ and the $R^2$ scores close to $0.99$, which implies that the predicted solution from the DNN matches the true very closely. We also note that the PDE solution predicted by the DNN is `smoother' than the FV solution of the PDE. This effects gets more pronounced when the lengthscales of the input diffusion field are smaller. The smoothness is a consequence of regularizing the DNN loss function.  
Fig. \ref{fig:errors_hist} shows the histograms of $\mathcal{E}$ and $R^2$ scores for all samples in $\calD_{\mathrm{test}}$. Note that all testing of the predictive capacity of the network is done using the test set $\calD_{\mathrm{test}}$ because the $\calD_{\mathrm{train}}$ and $\calD_{\mathrm{val}}$ have already been used during the training and model selection phase. 

\subsection{Predictions at arbitrary lengthscales}
\label{sec:diff_ls_exp}
\begin{figure}[h]
\subfigure[]{
\includegraphics[width=0.48\textwidth]{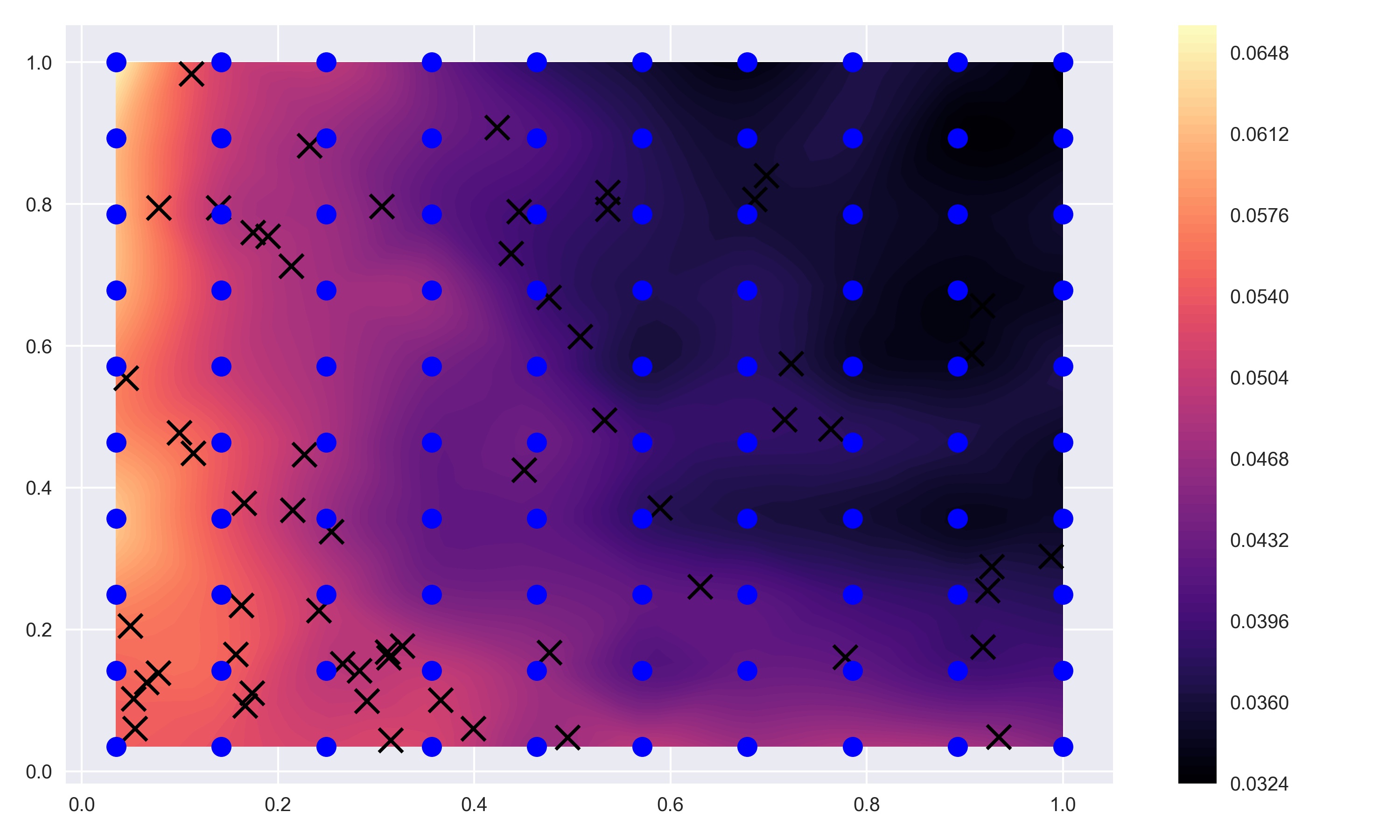}
\label{fig:arbit_exp_ls_relerrors}
}
\subfigure[]{
\includegraphics[width=0.48\textwidth]{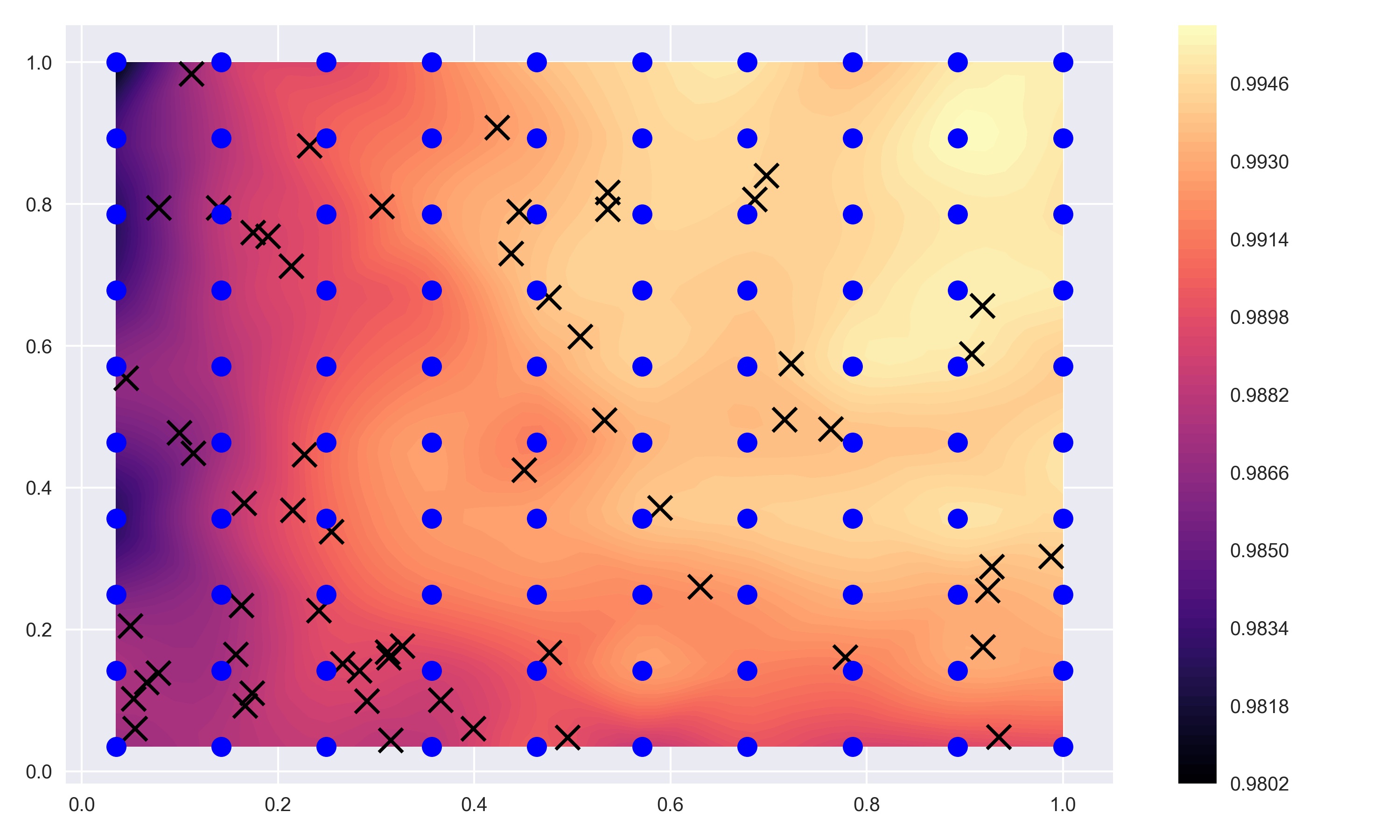}
\label{fig:arbit_exp_ls_r2scores}
}
\caption{\ref{fig:arbit_exp_ls_relerrors} - Mean relative errors of the predicted solution corresponding to samples of $a$ with arbitrary pairs of lengthscales not used in the DNN training. \ref{fig:arbit_exp_ls_r2scores} - Mean $R^2$ scores of the predicted solutions corresponding to samples of $a$ with arbitrary pairs of lengthscales not used in the DNN training. The 'x' markers correspond to lengthscales used in training the DNN and the solid dots correspond to lengthscales used to test the DNN surrogate.}
\label{fig:arbit_ls_exp_plots}
\end{figure}
Having constructed a DNN surrogate for the FV solution of the PDE, we would like to test predictive accuracy for samples of $\hat{\ax}$ with lengthscales which are not used to generate the dataset $\calD$. A $10 \times 10$ uniform grid of lengthscales is generated in the domain $[h, 1]^2$, and for each lengthscale, 100 observations of the diffusion field and it's corresponding PDE solution are generated. The mean of  the relative errors and mean of the $R^2$ scores for each lengthscale pair in this uniform grid is computed and shown in Fig. \ref{fig:arbit_ls_exp_plots}.
We observe that even when the input field has lengthscales that do not match the lengthscales used for training, we are able to predict the solution with accuracy similar to that obtained during testing with samples from $\calD_{\mathrm{test}}$. This suggests that DNN surrogate is learning to map the `picture' of the input field to the corresponding output. Note that the accuracy of the DNN decreases for diffusion fields with very fine lengthscales. This is consistent with the intuitive expectation that the less ``variation" in the structure of the diffusion field, the easier it is characterize the function that maps the input to the solution.

\subsection{Uncertainty Propagation}
\label{sec:spde_uq}

\begin{figure}[h]
\subfigure[]{
\includegraphics[width=0.5\textwidth]{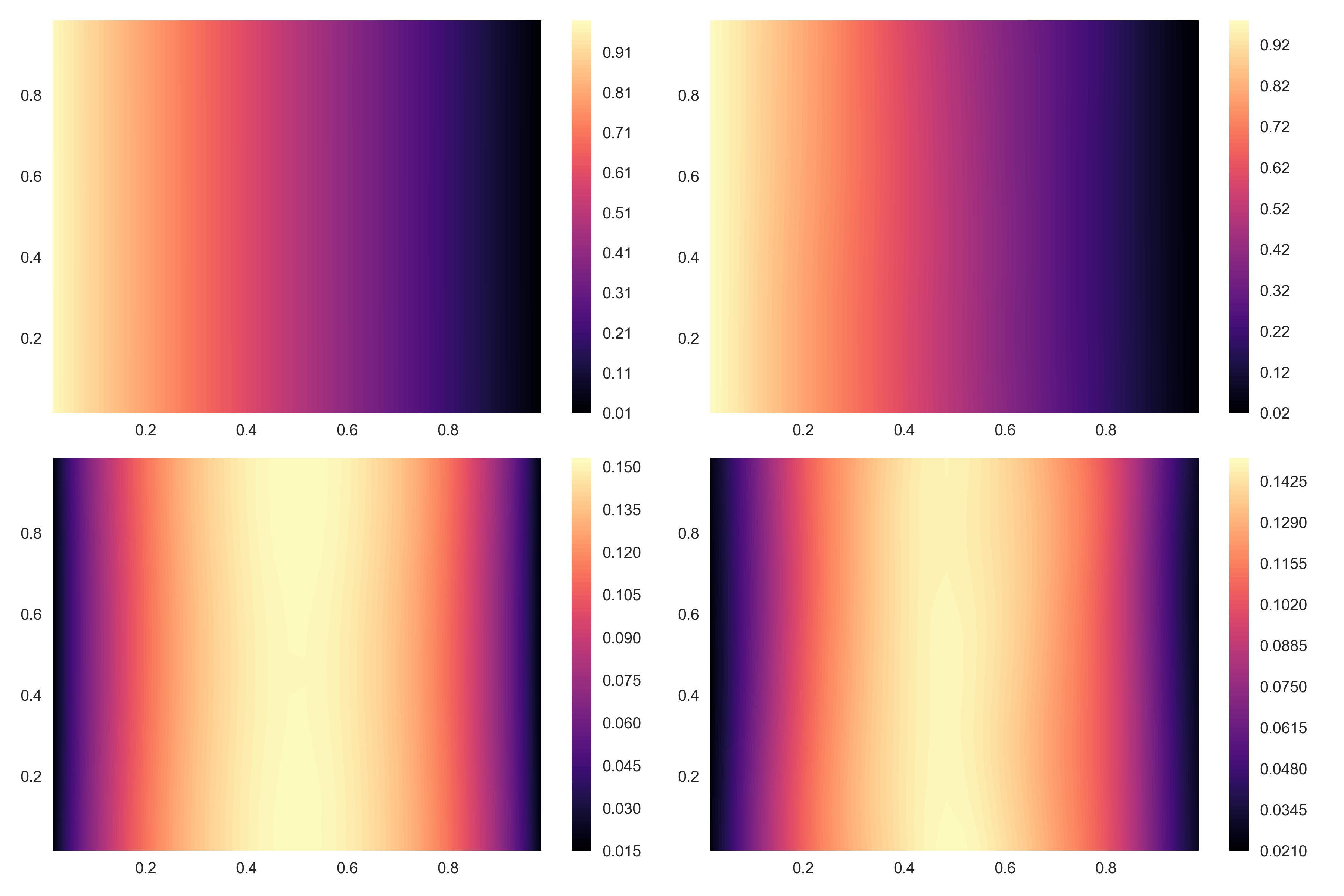}
\label{fig:up_case_1}
}
\subfigure[]{
\includegraphics[width=0.5\textwidth]{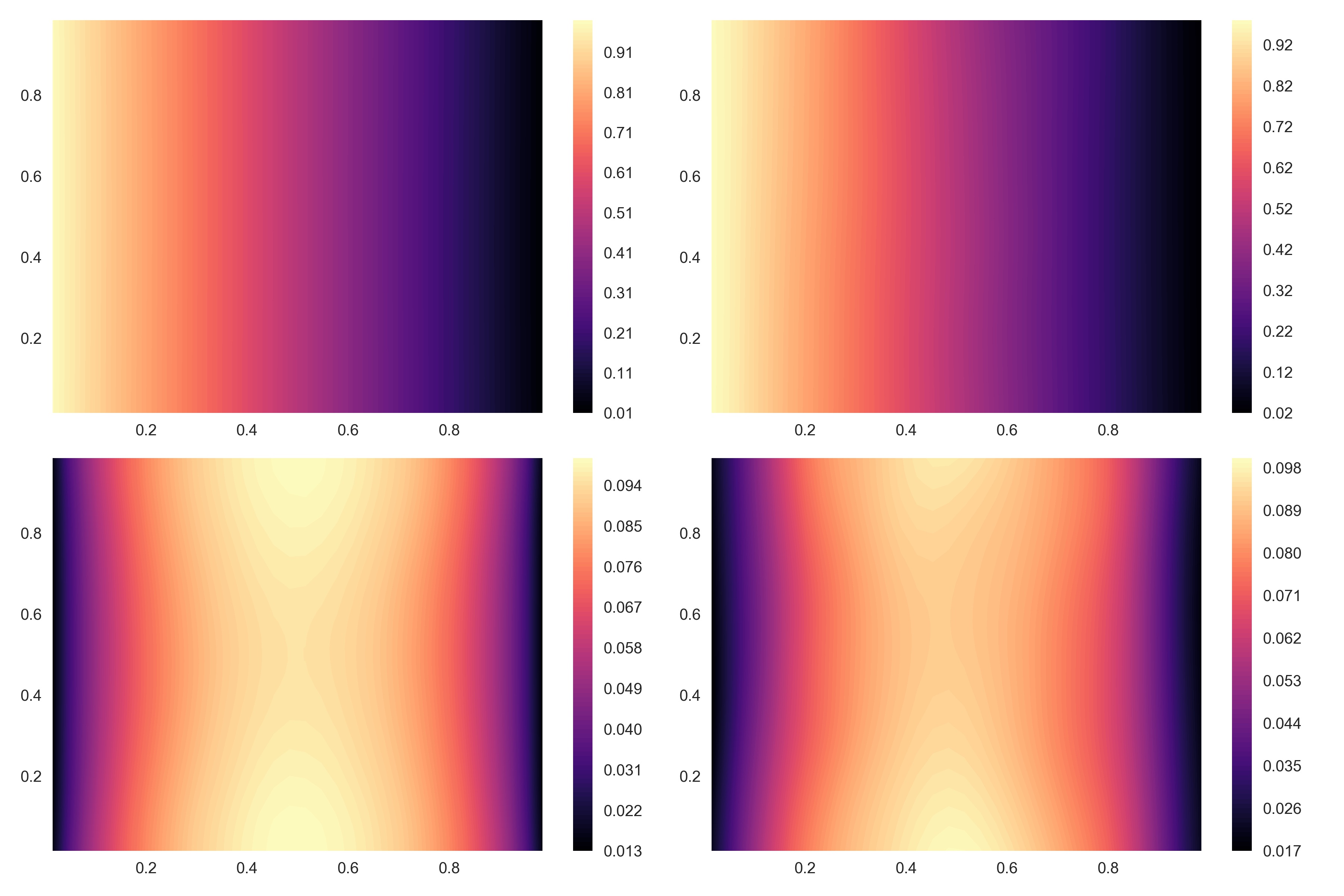}
\label{fig:up_case_2}
}
\caption{Mean and standard deviation of the PDE solution obtained by MC sampling of the DNN surrogate. In each sub figure the left column shows the MCS approximation and the right column shows the DNN approximation. The top half compares the mean of the solution and the bottom half compares the standard deviation. \ref{fig:up_case_1} - Case 1: $\ell_x = 0.1$ and $\ell_y = 0.5$. \ref{fig:up_case_2} - Case 2: $\ell_x = 0.05$ and $\ell_y = 0.15$. }
\label{fig:spde_uq}
\end{figure}

\begin{figure}[h]
\subfigure[]{
\includegraphics[width=0.5\textwidth]{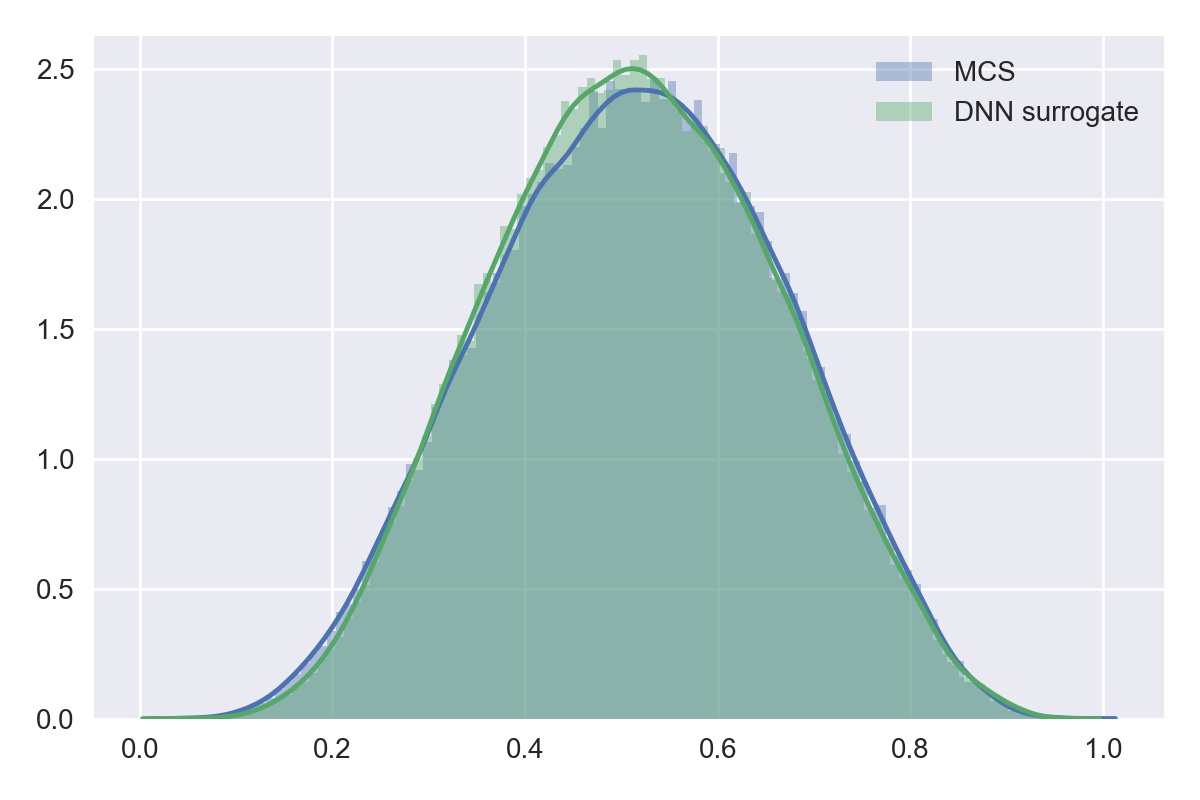}
\label{fig:up_case_1_pdf_1}
}
\subfigure[]{
\includegraphics[width=0.5\textwidth]{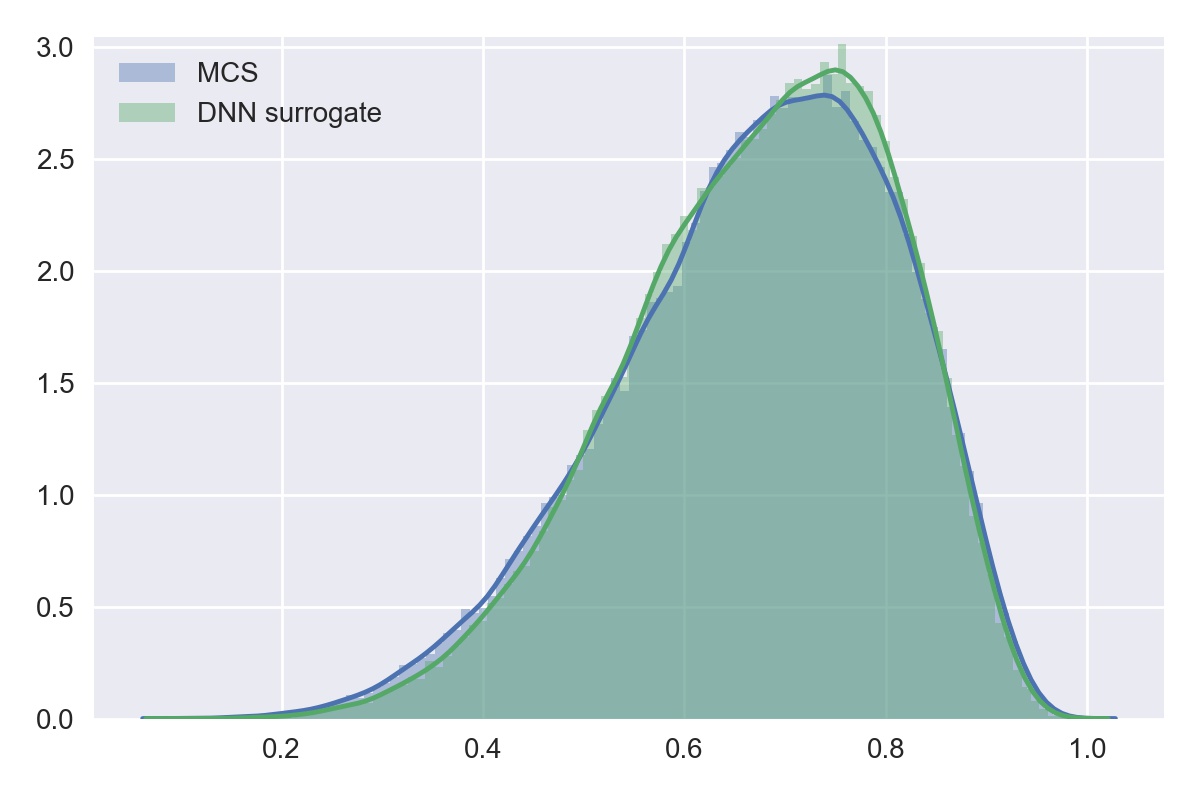}
\label{fig:up_case_1_pdf_2}
}
\subfigure[]{
\includegraphics[width=0.5\textwidth]{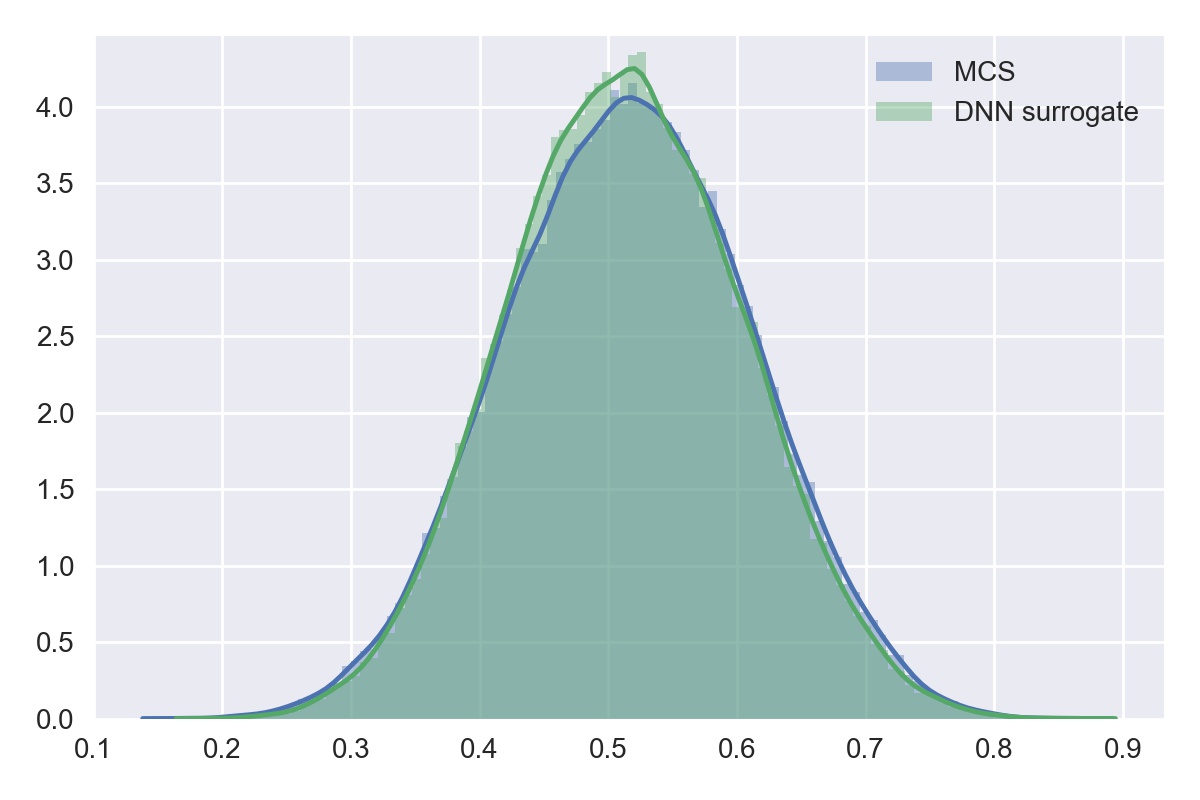}
\label{fig:up_case_2_pdf_1}
}
\subfigure[]{
\includegraphics[width=0.5\textwidth]{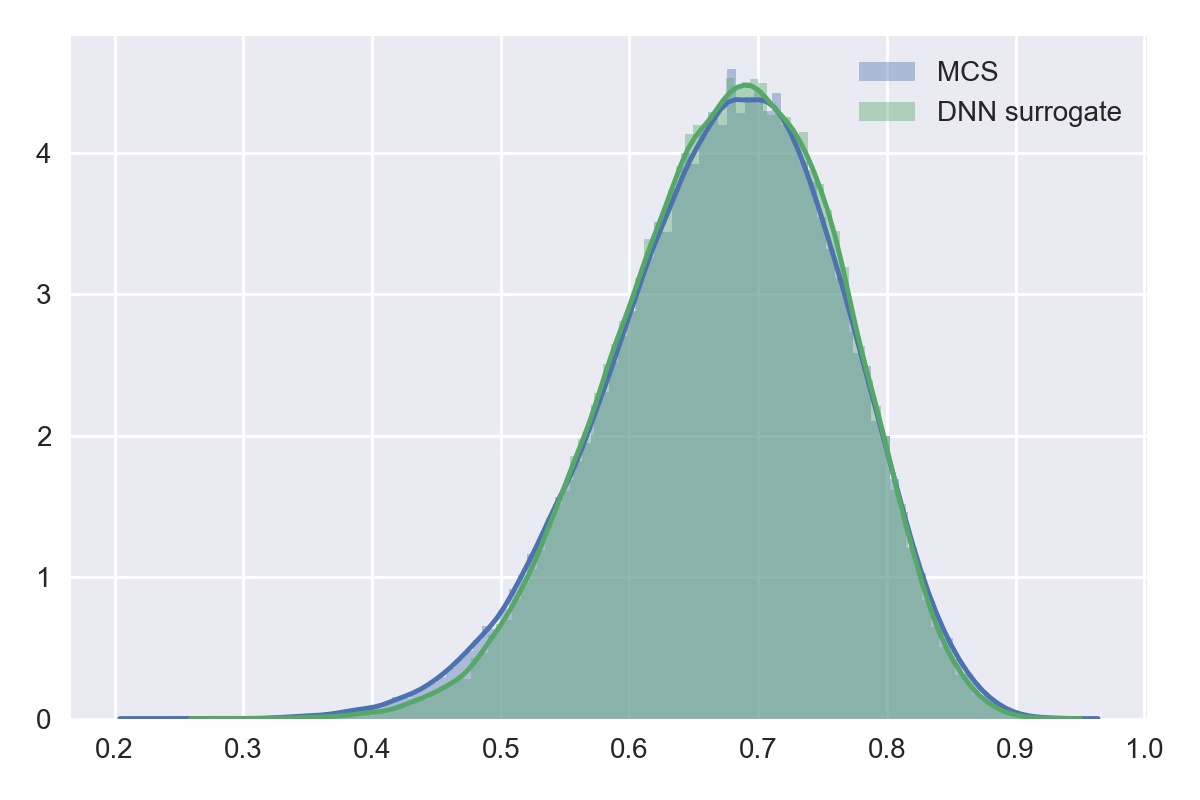}
\label{fig:up_case_2_pdf_2}
}
\caption{\ref{fig:up_case_1_pdf_1} and \ref{fig:up_case_1_pdf_2} - Density of PDE solution at $\x_1$ for cases 1 and 2 respectively. \ref{fig:up_case_2_pdf_1} and \ref{fig:up_case_2_pdf_2} - Density of PDE solution at $\x_2$ for cases 1 and 2 respectively.}
\label{fig:up_pdf}
\end{figure}
Having constructed a DNN surrogate that maps the input diffusion coefficient of the PDE in Eq. \ref{eqn:pde_bc} and verified its accuracy, we use this surrogate to solve a UP problem. This surrogate is generalizable for arbitrary choices of the lengthscale  of the input diffusion field. 
We consider the following 2 different choices of input lengthscales for propagating uncertainty -
\begin{enumerate}
\item Case 1: $\ell_x = 0.1$ and $\ell_y = 0.5$.
\item Case 2: $\ell_x = 0.05$ and $\ell_y = 0.15$. 
\end{enumerate}

In each case, we draw $10^{5}$ samples from the corresponding input distribution and estimate the following output statistics from the DNN surrogate predictions:
\begin{enumerate}
\item Mean of $\hat{\ux}$.
\item Variance of $\hat{\ux}$.
\item Probability density of the PDE solution at $\x_1 = (0.484,  0.484)$ and $\x_2 = (0.328, 0.641)$. 
\end{enumerate}
The comparison between DNN surrogate approximation of the above quantities and their corresponding MCS approximations, for cases 1 and 2, are shown in Fig. \ref{fig:spde_uq} and Fig. \ref{fig:up_pdf}. The relative error and $R^2$ scores between the DNN surrogate and the MCS approximations of  the mean and standard deviations are shown in Tab. \ref{tab:up_errors}.
\begin{table}[h!]
\centering
\begin{tabular}{||c | c | c | c | c||}
\hline & \multicolumn{2}{|c|}{Mean} & \multicolumn{2}{|c|}{Standard deviation} \\ \cline{2-5}
\textbf{Case} & $\mathcal{E}$ & $R^2$ & $\mathcal{E}$ & $R^2$ \\
\hline \hline
1 & 0.01174 & 0.99944 & 0.06565 & 0.96446 \\
\hline 
2 & 0.01080 & 0.99953 & 0.07035 & 0.95105 \\
\hline
\end{tabular}
\caption{Relative error and $R^2$ scores in the mean and variance of the PDE solution for two different choices of spatial lengthscale pairs. }
\label{tab:up_errors}
\end{table}
We note that the mean PDE solution from the DNN surrogate matches that from the MCS sampling very closely in both cases. The error in the standard deviation, while reasonably low, is increased because of the tendency of the DNN to `smooth out' the solution as discussed earlier. This is why we see a somewhat larger relative error for case 2, where the smaller lengthscales of the diffusion coefficient lead to PDE solutions that are inherently less smooth than the larger lengthscales of case 1.

%% file: conclusion.tex
\section{Conclusion}
\label{sec:conclusion}
We propose a methodology for learning DNN surrogate models for uncertainty quantification based on a parameterization of the DNN structure, such that the DNN is a composition of an encoder and one-layer MLP. Our parameterization lends the DNN surrogate the interpretation of recovering a nonlinear active subspace. We use a combination of grid search and BGO to select model hyperparameters, namely, the number of hidden layers, $L$, the width of the AS, $h$, and the weight decay regularization constant, $\lambda$. We demonstrate our methodology with a UP problem in elliptic SPDE with uncertain diffusion coefficient, and learn a surrogate which maps a `picture' of the discretized version of the coefficient to the PDE solution. Furthermore, we demonstrated that the DNN surrogate can effectively predict the solution of the PDE, even for diffusion fields with lengthscales that are not used for training the network.

This  work is an early step towards using deep learning to create surrogate models for high dimensional UQ tasks. UQ for state-of-the-art computational simulators are notoriously difficult because of the prohibitive time span for individual simulations. One can extend the methodology proposed in this work to a Bayesian treatment of DNNs\cite{blundell2015weight}, i.e., imposing a prior on the weights of the NN and using approximate inference techniques such as variational inference\cite{ranganath2014black, graves2011practical} to estimate the posterior distribution over the weights. Additionally, the Bayesian approach would allow one to better quantify the epistemic uncertainty induced by limited data. 

DNNs are also naturally suited for tasks for multilevel/multifidelity UQ \cite{peherstorfer2016survey, kennedy2000predicting, perdikaris2017nonlinear}. For instance, fully convolutional networks do not impose constraints on input dimensionality and can be trained on data obtained from several simulators at varying levels of fidelity. The hierarchical representation of information with a deep network can be used to learn correlations between heterogeneous information sources.